\documentclass[11pt,A4paper]{article}
\usepackage{amsmath,amssymb,bm,bbm,mathrsfs}
\usepackage[dvips]{graphicx}

\usepackage[all]{xy}
\usepackage{theorem}

\def\qed{\hfill \vrule height 7pt width 7pt depth 0pt
              \medskip}
\def\beq{\begin{equation}}
\def\eeq{\end{equation}}

\def\proof{\noindent{\bf Proof}\ \ }
\newtheorem{theorem}{Theorem}
\newtheorem{proposition}[theorem]{Proposition}
\newtheorem{lemma}[theorem]{Lemma}

\newtheorem{definition}{Definition}
\newtheorem{assumption}[definition]{Assumption}
{\theorembodyfont{\rmfamily}
}

\newcommand{\ds}{\displaystyle}

\newcommand{\ba}{\begin{array}}
\newcommand{\ea}{\end{array}}

\newcommand{\mb}{\boldsymbol}
\newcommand{\be}{\begin{equation}}
\newcommand{\ee}{\end{equation}}
\newcommand{\eps}{\varepsilon}
\newcommand{\ups}{\upsilon}
\newcommand{\mc}{\mathcal}

\newcommand{\ra}{\rightarrow}
\newcommand{\ov}{\overline}

\newcommand{\Z}{\mathbb{Z}}

\renewcommand{\1}{\mathbbm{1}}
\newcommand{\E}{\mathbb{E}}

\newcommand{\R}{\mathbb{R}}
\newcommand{\N}{\mathbb{N}}

\newcommand{\summ}{\sum\limits}
\renewcommand{\P}{\mathbb{P}}

\newcommand{\de}{\mathrm{d}}

\newcommand{\st}{\text{ s.t. }}

\DeclareMathOperator*{\argmax}{argmax}

\DeclareMathOperator{\ent}{H} 
\title{On the error exponent of variable-length block-coding schemes
over finite-state Markov channels with feedback
\thanks{
This work was partially supported by the National Science
Foundation under Grant ECCS-0547199. Earlier versions have been
presented at ISIT 2007, June 25-29 2007, Nice, France, and at ITW
2007, September 3-7 2007, Lake Tahoe, CA, USA. }}
\author{
Giacomo Como\thanks{Dipartimento di Matematica, Politecnico di
Torino, Corso Duca degli Abruzzi 24, 10126 Torino, Italy. Was with
Electrical Engineering, Yale University, New Haven, CT 06511, USA.
Email: {\tt\small giacomo.como@polito.it}}, Serdar
Y\"uksel\thanks{Mathematics and Engineering, Queen's University,
Kingston, Ontario, Canada, K7L 3N6 ON. Was with Electrical
Engineering, Yale University, New Haven, CT 06511, USA. Email:
{\tt\small serdar.yuksel@yale.edu}} and Sekhar Tatikonda
\thanks{Electrical Engineering, Yale University, 12 Hillhouse st,
New Haven, CT 06511, USA.
Email: {\tt\small sekhar.tatikonda@yale.edu}}%
} \date{}
\begin{document}
\maketitle
\begin{abstract}
The error exponent of Markov channels with feedback is studied
in the variable-length block-coding setting.
Burnashev's~\cite{Burnashev} classic result is extended and a
single letter characterization for the reliability function of
finite-state Markov channels is presented, under the assumption
that the channel state is causally observed both at the
transmitter and at the receiver side. Tools from stochastic
control theory are used in order to treat channels with
intersymbol interference. In particular the convex analytical
approach to Markov decision processes
\cite{BorkarConvexAnalytical} is adopted to handle problems with
stopping time horizons arising from variable-length coding
schemes.
\end{abstract}
\section{Introduction}
The role of feedback in channel coding is a long studied problem
in information theory. In 1956 Shannon \cite{Shannon} proved that
noiseless causal output feedback does not increase the capacity of
a discrete memoryless channel (DMC).
Feedback, though, can help in improving the trade-off between
reliability and delay of DMCs at rates below capacity. This
trade-off is traditionally measured in terms of error exponent; in
fact, since Shannon's work, much research has focused on studing
error exponents of channels with feedback. Burnashev
\cite{Burnashev} found a simple exact formula for the reliability
function (i.e. the highest achievable error exponent) of a DMC
with perfect causal output feedback in the variable-length
block-coding setting. The present paper deals with a
generalization of Burnashev's result to a certain class of
channels with memory. Specifically, we shall prove a simple
single-letter characterization of the reliability function of
finite-state Markov channels (FSMCs), in the general case when
intersymbol-interference (ISI) is present. Under mild ergodicity
assumptions, we will prove that, when one is allowed
variable-length block-coding with perfect causal output feedback
and causal state knowledge both at the transmitter and at the
receiver end, the reliability function has the form \be E_B(R) = D
\left( 1 - \frac{R}{C}\right)\,,\qquad R\in(0,C)\label{E_B}\,. \ee
In (\ref{E_B}), $R$ is the transmission rate, measured with
respect to the average transmission time. The capacity $C$ and the
coefficient $D$ are quantities which will be defined as solution
of finite dimensional optimization problems involving the
stochastic kernel describing the FSMC. The former will turn out to
equal the maximum, over all choices of the channel input
distributions as a function of the channel state, of the conditional
mutual information between channel input and the pair of channel output
and next channel state given the current state, whose marginal distribution
coincides with the induced ergodic state measure (see (\ref{capacitydef})).
The latter will instead equal the average, with respect to the induced
ergodic state measure, of the
Kullback-Leibler information divergence between the joint channel
output and next channel state distributions associated to the pair of
most distinguishable choices of a channel input symbol as a function of
the current state (see (\ref{Burnashevcoeffdef})).

The problem of characterizing error exponents of memoryless
channels with feedback has been addressed in the information
theory literature in a variety of different frameworks.
Particularly relevant are the choice of block versus continuous
transmission, the possibility of allowing variable-length coding
schemes, and the way delay is measured. In fact, much more than in
the non-feedback case, these choices lead to very different
results for the error exponent, albeit not altering the capacity
value. In continuous transmission systems information bits are
introduced at the encoder, and later decoded, individually.
Continuous transmission with feedback was considered by Horstein
\cite{Horstein}, who was probably the first showing that
variable-length coding schemes can give larger error exponents
than fixed-length ones.
Recently, continuous transmission with fixed delay has attracted
renewed attention in the context of anytime capacity \cite{Sahai}.
In this paper, however, we shall restrict ourselves to block
transmission, which is the framework considered by the largest
part of the previous literature.

In block transmission systems the information sequence is
partitioned into blocks of fixed length which are then encoded
into channel input sequences. When there is no feedback these
sequences need to be of a predetermined, fixed length in order to
guarantee that transmitter and receiver remain synchronized. When
there is feedback, instead, the availability of common information
shared between transmitter and receiver makes it possible to use
variable-length schemes. Here the transmission time is allowed to
dynamically depend on the channel output sequence. It is known
that exploiting the possibility of using variable-length
block-coding schemes guarantees high gains in terms of error
exponent. In fact, Dobrushin \cite{Dobrushin} showed that the
sphere-packing bound still holds for fixed-length block-coding
schemes over symmetric DMCs even when perfect output feedback is
causally available the encoder (a generalization to nonsymmetric
DMCs was addressed in \cite{Haroutunian}). Even though
fixed-length block-coding schemes with feedback have been studied
(see \cite{Zigangirov,Djackov}) the above-mentioned results pose
severe constraints on the performance such schemes can achieve.
Moreover, no closed form for the reliability function at all rates
is known for fixed-length block coding with feedback, but for the
very special class of symmetric DMCs with positive zero-error
capacity (cf. \cite[pag.199]{Csiszarbook}). It is worth to mention
that the situation can be much different for continuous alphabet
channels. For the additive white Gaussian noise channel (AWGNC)
with average power constraint, Shalkwijk and Kailath
\cite{SchalkwijkKailath} proved that a doubly exponential error
rate is achievable by fixed-length block codes. However, when a
peak power constraint to the input of an AWGNC is added, then this
phenomenon disappear as shown in \cite{Wyner}. At the same time it
has been also well-known that, if variable length coding schemes
are allowed, then the sphere-packing exponent can be beaten even
when no output feedback is available but for a single una tantum
bit guaranteeing synchronization between transmitter and receiver.
This situation is traditionally referred to as decision feedback
and was studied in \cite{Forney} (see also
\cite[pag.201]{Csiszarbook}).

A very simple exact formula was found by Burnashev
\cite{Burnashev} for the reliability function of DMCs with full
causal output feedback in the case variable-length block-coding
schemes. Burnashev's analysis combined martingale theory arguments
with more standard information theoretic tools. It is remarkable
that in this setting the reliability function is known, in a very
simple form, at any rate below capacity, in sharp contrast to what
happens in most channel coding problems for which the reliability
function can be exactly evaluated only at rates close to capacity.
Another important point is that Burnashev exponent of a generic
DMC can dramatically exceed the sphere-packing exponent; in
particular it approaches capacity with nonzero slope.

Thus, variable-length block coding appears a natural setting for
transmission over channels with feedback. In fact,
it has already been considered by many authors after Burnashev's
landmark work. A simple two-phase iterative scheme achieving
Burnashev exponent was introduced by Yamamoto and Itoh in
\cite{YamamotoItoh}. More recently, low-complexity variable-length
block-coding schemes with feedback have been proposed and analyzed
in \cite{OoiWornell}. The works \cite{TchamkertenTelatar1} and
\cite{TchamkertenTelatar2} dealt with universality issues,
addressing the question whether Burnashev exponent can be achieved
without exact knowledge of the statistics of the channel but only
knowing it belongs to a certain class of DMCs. In
\cite{TelatarRimoldi} a simplification of Burnashev's original
proof \cite{Burnashev} is proposed, while \cite{GallagerNakiboglu}
is concerned with the characterization of the reliability function
of DMCs with feedback and cost constraints. In \cite{Ooithesis}
low-complexity schemes for FSMCs with feedback are proposed.
However, to the best of our knowledge, no extension of Burnashev's
theorem to channels with memory has been considered.

The present work deals with a generalization of Burnashev's result
to FSMCs. As an example, channels with memory, and FSMCs in
particular, model transmission problems where fading is an
important component as for instance in wireless communication.
Information theoretical limits of FSMCs both with and without
feedback have been widely studied in the literature: we refer to
the classic textbooks \cite{Gallagerbook,Wolfowitz} and references
therein for overview of the available literature (see also \cite{GoldsmithVaraya}).
It is known that the capacity is strongly affected by the hypothesis about the
nature of the channel state information (CSI) both available at
the transmitter and at the receiver side. In particular while
output feedback does not increase the capacity when the state is
causally observable both at the transmitter and at the receiver
side (see \cite{TatikondaMitter} for a proof, first noted in \cite{Shannon}),
it generally does so for different information patterns. In particular,
when the channel state is not observable at the transmitter, it is known that
feedback may help
improving capacity by allowing the encoder in estimating the
channel state \cite{TatikondaMitter}. However, in this paper only
the case when the channel state is causally observed both at the
transmitter and at the receiver end will be considered. Our choice
is justified by the aim to separate the study of the role of
output feedback in channel state estimation from its effect in
allowing better reliability versus delay tradeoffs for
variable-length block-coding schemes.

In \cite{TatikondaMitter} a general stochastic control framework
for evaluating the capacity of channels with memory and feedback
has been introduced. The capacity has been characterized as the
solution of a dynamic programming average cost optimality
equation. Existence of a solution to such an equation implies
information stability. Also lower bounds \`a la Gallager to the
error exponents achievable with fixed-length coding schemes are
obtained in \cite{TatikondaMitter}. In the present paper we follow
a similar approach in order to characterize the reliability
function of variable-length block-coding schemes with feedback.
Such an exponent will be characterized in terms of solutions to
certain Markov decision problems. The main new feature posed by
variable-length schemes is that we have to deal with average cost
optimality problems with a stopping time horizon, for which
standard results in Markov decision theory cannot be used
directly. We adopt the convex analytical approach of
\cite{BorkarConvexAnalytical} and use Hoeffding-Azuma inequality
in order to prove a strong uniform convergence result for the
empirical measure process. This allows us to find sufficient
conditions on the tails of a sequence of stopping times for the
solutions of the corresponding average cost optimality problems to
be asymptotically approximated by the solution of the corresponding
infinite horizon problem, for which stationary policies are known to be optimal.

The rest of this paper is organized as follows. In Section 2
causal feedback variable-length block-coding schemes for FSMCs are
introduced, and capacity and reliability function are defined as
solution of optimization problems involving the stochastic kernel
describing the FSMC. The main result of the paper is then stated
in Theorem \ref{maintheo}. In Section 3 we prove an upper
bound to the best error exponent achievable by variable-length
block-coding schemes with perfect feedback over FSMCs.
The main result of that section is contained in Theorem \ref{invtheo}
which generalizes Burnashev result.
Section 4 is of a technical nature and
deals with Markov decision processes with stopping time horizons.
Some stochastic control techniques are reviewed and the main result
is contained in Theorem \ref{lemmac*} which is then
used to prove that the bound of Theorem \ref{invtheo}
asymptotically coincides with the reliability function (\ref{E_B}).
In Section 5 a family of simple
iterative schemes based on a generalization of Yamamoto-Itoh's
\cite{YamamotoItoh} is proposed and its performance is analyzed
showing that this family is asymptotically optimal in terms of
error exponent. Finally, in Section 6 an explicit example is studied.
Section 7 presents some conclusions and
points out to possible topics for future research.

\section{Statement of the problem and main result}
\label{sect2}
\subsection{Stationary ergodic Markov channels}
\label{sect2.1}
Throughout the paper $\mc X$, $\mc Y$, $\mc S$ will respectively
denote channel input, output and state spaces. All are assumed to
be finite.
\begin{definition}\label{defchannel}
A stationary \emph{Markov channel} is described by:
\begin{itemize}
    \item a stochastic kernel consisting in
          a family $\left\{P(\,\cdot\,,\,\cdot\,|\,s,x)
          \in\mc P(\mc S\times\mc Y)|s\in\mc S,x\in\mc X\right\}$
          of probability measures over $\mc S\times\mc Y$,
          indexed by elements of $\mc S$ and $\mc X$;
    \item an initial state distribution $\mc{\mb\mu}_1$ in $\mc P(\mc S)$.
\end{itemize}
\end{definition}
For a channel as in Def.\ref{defchannel}, let
$$P_{S}(s_+|\,s,x):=\sum_{y\in\mc Y}P(s_+,y|\,s,x)$$
be the $\mc S$-marginals. We shall say that a Markov channel as
above has no ISI when the $\mc S$-marginals do not depend on the
chosen channel input, i.e. \be P_{\mc S}(s_+|\,s,x_1)=P_{\mc
S}(s_+|\,s,x_2)\,,\qquad\forall\,s,s_+\in\mc S\,, \ x_1,x_2\in\mc
X\,.\label{noISI}\ee We will consider the associated stochastic
kernels
$$
\left\{Q(\,\cdot\,,\,\cdot\,|\,s,\mb u)\in\mc P(\mc S\times\mc
Y)|\,s\in\mc S,\mb u\in\mc P(\mc X)\right\}\,,\qquad
\left\{Q_{S}(\,\cdot\,|\,s,\mb u)\in\mc P(\mc S)|\,s\in\mc S,\mb
u\in\mc P(\mc X)\right\}\,,$$ where for every channel input
distribution $\mb u$ in $\mc P(\mc X)$ \be\label{Qdef}
Q(s_+,y|\,s,\mb u):=\sum_{x\in\mc X}P(s_+,y\,|\,s,x)\mb
u(x)\,,\qquad Q_S(s_+|\,s,\mb u):=\sum_{x\in\mc
X}P_S(s_+\,|\,s,x)\mb u(x)\,.\ee

Given $\pi:\mc S\ra\mc P(\mc X)$ (we shall refer to such a map as
a deterministic stationary policy), denote by \be\label{Qpidef}
Q_{\mb\pi}:=\left(Q\big(s_+\,|\,s,\pi(s)\big)\right)_{s,s_+\in\mc
S}\ee the state transition stochastic matrix induced by $\pi$.
With an abuse of notation, for any map $f:\mc S\ra\mc X$ we shall
write $Q_{f}$ in place of $Q_{\delta_{f(\cdot)}}$. Throughout the
paper we will restrict ourselves to FSMCs satisfying the following
ergodicity assumption.
\begin{assumption}
For every $f:\mc S\ra\mc X$ the stochastic matrix $Q_{f}$ is
irreducible. \label{mixingassumption}
\end{assumption}
Assumption \ref{mixingassumption} can be relaxed or replaced by
other equivalent assumptions. Here we limit ourselves to observe
that it involves the $S$-marginals $\{P_{S}\}$ of the
Markov channel only. Moreover it is purely discrete condition, since
it requires a finite number of finite directed graphs to be
strongly connected. Since taking a convex combination does not
reduce the support, Assumption \ref{mixingassumption} guarantees
that for every deterministic stationary policy $\pi:\mc S\ra\mc
P(\mc X)$ the stochastic matrix $Q_{\pi}$ is irreducible. Then,
Perron-Frobenius theorem guarantees that $Q_{\pi}$ has a unique
invariant measure in $\mc P(\mc S)$ which will be denoted by
$\mb\mu_{\pi}$. Notice that in the non-ISI case Assumption
\ref{mixingassumption} is tantamount to requiring the strict
positivity of the $\mc S$-marginals of the stochastic kernel.

\subsection{Capacity of ergodic FSMCs}
To any ergodic FSMC we associate the mutual information cost function
$c:\mc S\times\mc P(\mc X)\ra\R$, \be c(s,\mb u)=\summ_{x\in\mc
X}\summ_{y\in\mc Y}\summ_{v\in\mc S} \mb
u(x)P(v,y|\,s,x)\log\frac{P(v,y|\,s,x)}{\summ_{z\in\mc X}\mb
u(z)P(v,y|\,s,z)}\,, \label{cdef}\ee and define its capacity as
\be C:=\max\limits_{\pi:\mc S\ra\mc P(\mc X)} \summ_{s\in\mc
S}\mb\mu_{\pi}(s)c(s,\pi(s))= \max\limits_{\pi:\mc S\rightarrow\mc
P(\mc X)}I(X;Y,S_+|\,S) \,.\label{capacitydef} \ee In the
rightmost side of (\ref{capacitydef}) the term $I(X;S_+,Y|\,S)$
denotes the conditional mutual information \cite{CoverThomas}
between $X$ and the pair $(S_+,Y)$ given $S$, where $S$ is an $\mc
S$-valued r.v. whose marginal distribution is given by the
invariant measure $\mb\mu_{\pi}$, $X$ is an $\mc X$-valued r.v.
whose conditional distribution given $S$ is described by the
policy $\pi$, while $S_+$ and $Y$ are respectively an $\mc
S$-valued r.v. and a $\mc Y$-valued r.v. whose joint conditional
distribution given $X$ and $S$ is described by the stochastic
kernel $P(S_+,Y|\,S,X)$. Notice that in particular the mutual
information cost function $c$ is continuous over $\mc S\times\mc P(\mc X)$
and takes values in the bounded interval $[0,\log|\mc X|]$ .

The quantity $C$ defined above is known to equal the capacity of
the ergodic Markov channel we are considering when perfect causal
CSI is available at both transmission ends, with or without output
feedback~\cite{TatikondaMitter}.
It is important to observe that, due to the presence of ISI in the
channel model we are considering, the policy $\pi$ plays a dual
role in the optimization problem in (\ref{cdef}) since it affects
both the mutual information cost $c(s,\pi(s))=I(X;S_+,Y|S=s)$ and
the ergodic channel state distribution $\mb\mu_{\pi}$ with respect
to which the former is averaged.

In the case when there is no ISI, i.e. when (\ref{noISI}) is
satisfied, this phenomenon disappears. In fact, since the
invariant measure $\mb\mu$ is independent of the policy $\pi$ we
have that (\ref{capacitydef}) reduces to \be C=\summ_{s\in\mc
S}\mb\mu(s)\max_{p_X\in\mc P(\mc X)}c(s,p_X)= \summ_{s\in\mc
S}\mb\mu(s)\max\limits_{p_X\in\mc P(\mc
X)}I(X;Y|S=s)\,,\label{CnoISI}\ee where in the rightmost side of
(\ref{CnoISI}) the quantity $\max_{p_X\in\mc P(\mc X)}I(X;Y|S=s)$
coincides with the capacity of the DMC associated to the state
$s$. The simplest case of FSMCs with no ISI is obtained when the
state sequence forms an i.i.d. process independent from the
channel input with distribution $\mb\mu$, i.e. when
$$P_{\mc S}(s_+|\,s,x)=\mb\mu(s_+)\,,\qquad \forall s,s_+\in\mc S\,,x\in\mc X\,.$$
In this case, it is not difficult to check that
(\ref{capacitydef}) reduces to the capacity of a DMC with input
space $\mc X'=\mc S^{\mc X}$ -the set of all maps from $\mc S$ to
$\mc X$-, output space $\mc Y':=\mc S\times\mc Y$ -the Cartesian
product of $\mc S$ times $\mc Y$-, and transition probabilities
given by \be P'(y'|\,x'):=\summ_{s\in\mc
S}\mb\mu(s)P(y'|\,s,x'(s))\,, \qquad x':\mc S\rightarrow\mc X\,,\
y'\in\mc S\times\mc Y\,.\label{equivchannel}\ee Observe the
difference with respect to the case when the state is causally
observed at the transmitter only, whose capacity was first found
in \cite{ShannonSideInfo}. While the input space of the equivalent
DMC is the same in both cases, its output space is larger in the
case we are dealing with in this paper with respect to that
addressed by Shannon, since we are assuming that the state is
causally observable also at the receiver end.

Finally, notice that, when the state space is trivial (i.e. when
$|\mc S|=1$), (\ref{capacitydef}) reduces to the usual definition
of the capacity of a DMC.

\subsection{Burnashev coefficient of FSMCs}
Consider now the cost function $d:\mc S\times\mc P(\mc
X)\ra[0,+\infty]$ \be d(s,\mb u):=\sup\limits_{\mb u'\in\mc P(\mc
X)} \summ_{y\in\mc Y}\summ_{s_+\in\mc S} \mb
u(x)Q(s_+,y|s,\mb u)\log\frac {Q(s_+,y|s,\mb u)}{Q(s_+,y|s,\mb u')}\,. \label{ddef}\ee
Notice that the term to be optimized in the righthand side of
(\ref{ddef}) equals the Kullback-Leibler information divergence
between the probability measures $Q(\,\cdot\,,\,\cdot\,|\,s,\mb u)$ and
$Q(\,\cdot\,,\,\cdot\,|\,s,\mb u')$ in $\mc P(\mc S\times\mc Y)$.
It follows that, if we introduce the quantities
\be\lambda:=\min\{\lambda_s|\,s\in\mc S\}\,,\qquad
\lambda_s:=\min\Big\{\min_{x\in\mc X}P(s_+,y|\,s,x)\big|\,
s_+,y:\exists\, z:\,P(s_+,y|\,s,z)>0\Big\}\,,\label{lambdadef}\ee
we have that the cost function $d$ is bounded and continuous over
$\mc S\times\mc P(\mc X)$ if and only if $\lambda$ is strictly
positive, i.e. \be\lambda>0\qquad\Longleftrightarrow\qquad
d_{\max}:=\sup\limits_{s\in\mc S,\mb u\in\mc P(\mc X)}d(s,\mb
u)<+\infty\,.\label{lambda>0}\ee Define the Burnashev coefficient
of a Markov channel as \be D:=\sup_{\pi:\mc S\ra\mc P(\mc
X)}\summ_{s\in\mc S}\mb\mu_{\pi}(s)d(s,\pi(s))\,.
\label{Burnashevcoeffdef}\ee

Notice that $D$ is finite iff (\ref{lambda>0}) holds. Moreover, a
standard convexity argument allows to conclude that both the
suprema in (\ref{ddef}) and in (\ref{Burnashevcoeffdef}) are
achieved in some corner points, so that \be \ba{rcl}\label{Dx0x1}
D&=&\max\limits_{f_0,f_1:\mc S\ra\mc X} \summ_{s\in\mc
S}\mb\mu_{f_0}(s)\summ_{s_+\in\mc S}\summ_{y\in\mc Y}
P(s_+,y|\,s,f_0(s))\log\ds\frac{P(s_+,y|\,s,f_0(s))}{P(s_+,y|\,s,f_1(s))}\\[15pt]
&=&\max\limits_{f_0,f_1:\mc S\ra\mc X}\summ_{s\in\mc
S}\mb\mu_{f_0}(s)
D\left(P(\,\cdot\,,\,\cdot\,|\,s,f_0(s))||\,P(\,\cdot\,,\,\cdot\,|\,s,f_1(s))\right)\,.\ea
\ee Similarly to what already noted for the role of policy $\pi$
in the optimization problem (\ref{capacitydef}), it can be
observed that, due to the presence of ISI, the map $f_0$ has a
dual effect in the maximization in (\ref{Dx0x1}) since it affects
both the Kullback-Leibler information divergence cost
$D\left(P(\,\cdot\,,\,\cdot\,|\,s,f_0(s))||\,P(\,\cdot\,,\,\cdot\,|\,s,f_1(s))\right)$
and the ergodic state measure $\mb\mu_{f_0}$. Notice the asymmetry
with the role of the map $f_1$ whose associated ergodic measure
instead does not come into the picture at all in the definition of the coefficient $D$.
Once again, in the absence of ISI, (\ref{Dx0x1}) simplifies to
$$D=\summ_{s\in\mc S}\mb\mu(s)\max\limits_{x_0,x_1\in\mc X}
D\left(P(\,\cdot\,,\,\cdot\,|\,s,x_0)||\,P(\,\cdot\,,\,\cdot\,|\,s,x_1)\right)\,.$$

We observe that in the memoryless case (which can be recovered
when $|\mc S|=1$) the coefficient $D$ coincides with the
Kullback-Leibler information divergence between the output
measures associated to the pair of most distinguishable inputs,
the quantity originally denoted with the symbol $C_1$ in
\cite{Burnashev}. When the state space is nontrivial ($|\mc
S|>1$), and the channel state process forms an i.i.d. sequence
independent from the channel input, then the Burnashev coefficient
$D$ reduces to that of the equivalent DMC with enlarged input
space $\mc X'=\mc X^{\mc S}$ and output space $\mc Y'=\mc
S\times\mc Y$ with transition probabilities defined in
(\ref{equivchannel}).


\subsection{Causal feedback encoders, sequential decoders, and main result}
\begin{figure}
\centering
\includegraphics[height=3.0cm,width=12cm]{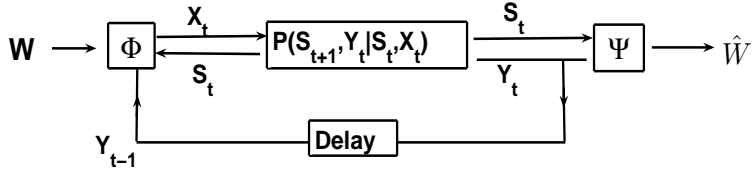}
\caption{Information patterns for variable-length block-coding schemes
on a FSMC with causal feedback and CSI.} \label{figurecodingsetting}
\end{figure}
\begin{definition}\label{defencoder}
A \emph{causal feedback encoder} is the pair of a finite message
set and a sequence of maps \be\Phi=\left(\mc W,\left\{\phi_t:\mc
W\times\mc Y^{t-1}\times\mc S^t\ra\mc X\right\}_{t\in\N}\right)\,.
\label{causalencdef} \ee
\end{definition}
With Def.\ref{defencoder}, we are implicitly assuming that perfect
state knowledge as well as perfect output feedback are available
at the encoder side.

Given a stationary Markov channel and a causal feedback encoder as
in Def.\ref{defencoder}, we will consider a probability space
$(\Omega,\mc A,\P_{\Phi})$ ($\E_{\Phi}$ will denote the
corresponding expectation operator) over which are defined:
\begin{itemize}
    \item a $\mc W$-valued random variable $W$ describing the message
          to be transmitted;
    \item a sequence $\mb X=(X_t)_{t\in\N}$ of $\mc X$-valued r.v.s
          (the channel input sequence);
    \item a sequence  $\mb Y=(Y_t)_{t\in\N}$ of
          $\mc Y$-valued r.v.s (the channel output sequence);
    \item a sequence $\mb S=(S_t)_{t\in\N}$ of $\mc S$-valued r.v.s (the state sequence).
\end{itemize}
We shall consider the time ordering
$$W,S_1,X_1,Y_1,S_2,X_2,Y_2,\ldots\,,$$
and assume that
$$\P_{\Phi}(W=w)=\frac1{|\mc W|}\,,\qquad
\P_{\Phi}(S_1=s\big|W)=\mb\mu(s)\,,$$
$$\P_{\Phi}(X_t=x\big|\,W,\mb S_1^t,\mb X_1^{t-1},\mb Y_1^{t-1})=
\delta_{\left\{\phi_t\left(W,\mb Y_1^{t-1},\mb
S_1^t\right)\right\}}(x)\,,\qquad \P_{\Phi}-a.s.\,,$$
$$\P_{\Phi}(S_{t+1}=s,Y_{t}=x\big|\,W,\mb S_1^t,\mb Y_1^{t-1},\mb X_1^t)=
P\left(s,y\big|S_t,X_t\right)\,,\qquad\P_{\Phi}-a.s.\,.$$

It is convenient to introduce the following notation for the
information patterns available at the encoder and decoder side.
For every $t$ we define the sigma-fields $\mc E_t:=\sigma\left(\mb
S_1^{t},\mb Y_1^{t-1}\right)$, describing the feedback information
available at the encoder side, and $\mc F_t:=\sigma\left(\mb
S_1^t,\mb Y_1^{t}\right)$, describing the information available at
the decoder. Clearly \be\{\emptyset,\Omega\}\!=\!\mc E_0\!=\!\mc
F_0\!\subseteq\!\mc E_1\!\subseteq \!\mc
F_1\!\subseteq\!\ldots\subseteq\mc A\,.\label{nestedfiltration}\ee
In particular we end up with two nested filtrations: \mbox{$\mc
F:=(\mc F_t)_{t\in\Z^+}$} and \mbox{$\mc E:=(\mc
E_{t})_{t\in\Z_+}$}.

\begin{definition}\label{defdectime}
A transmission time $T$ is a stopping time for the filtration $\mc
F$.
\end{definition}
\begin{definition}\label{defdecoder}
Given a causal feedback encoder $\Phi$ as in (\ref{causalencdef})
and a transmission time $T$, a sequential decoder for $\Phi$ and
$T$ is a $\mc W$-valued $\mc F_T$-measurable random variable.
\end{definition}
Notice that with Def.s \ref{defdectime} and \ref{defdecoder} we
are assuming that perfect causal state knowledge is available at
the receiver. In particular the fact that the transmitter's and
the receiver's information patterns are nested guarantees that
encoder and decoder stay synchronized while using a
variable-length scheme.

Given a causal feedback encoder $\Phi$ as in Def.
\ref{defencoder}, a transmission time $T$ and a sequential decoder
$\Psi$, we will call the triple $(\Phi,T,\Psi)$ a variable-length
block-coding scheme and define its error probability as
$$p_e(\Phi,T,\Psi):=\P_{\Phi}\left(\Psi\ne W\right)\,.$$
Following Burnashev's approach we shall consider the expected
decoding time $\E_{\Phi}[T]$ as a measure of the delay of the
scheme $(\Phi,T,\Psi)$ and accordingly define its rate as
$$R(\Phi,T,\Psi):=\frac{\log|\mc W|}{\E_{\Phi}[T]}\,.$$

We are now ready to state our main result. It is formulated in an
asymptotic setting, considering countable families of causal
encoders and sequential decoders with asymptotic average rate
below capacity and vanishing error probability.
\begin{theorem}
For any $R$ in $(0,C)$
\begin{enumerate}
    \item  any family $(\Phi^n,T_n,\Psi^n)_{n\in\N}$
         of variable-length block-coding schemes such that
\be\lim\limits_{n\in\N}p_e(\Phi^n,T_n,\Psi^n)=0\,, \qquad
\liminf\limits_{n\in\N}R(\Phi^n,T_n,\Psi^n)\ge
R\,,\label{constr}\ee satisfies
\be\limsup\limits_{n\in\N}-\frac1{\E_{\Phi^n}[T_n]} \log
p_e\left(\Phi^n,T_n,\Psi^n\right)\le E_B(R)\,.\label{maintheo1}\ee
    \item
there exists a family $\left(\Phi^n,T_n,\Psi^n\right)_{n\in\N}$ of
variable-length block-coding schemes satisfying (\ref{constr}) and
such that
\begin{itemize}
\item if $D<+\infty$ \be \lim\limits_{n\in\N}
-\frac1{\E^{\Phi^n}[T_n]}\log p_e\left(\Phi^n,T_n,\Psi^n\right)=
E_B(R)\,, \label{maintheo2} \ee \item if $D=+\infty$ \be
p_e\left(\Phi^n,T_n,\Psi^n\right)=0\,,\qquad \forall
n\in\N\,.\label{zeroerror}\ee
\end{itemize}
\end{enumerate}
\label{maintheo}
\end{theorem}
We observe that Burnashev's original result \cite{Burnashev} for
memoryless channels can be recovered as a particular case of
Theorem \ref{maintheo} when the state space is trivial, i.e.
$|\mc S|=1$.

\section{An upper bound on the achievable error exponent }
\label{sect4} The aim of this section is to provide an upper bound
on the error exponent of an arbitrary variable-length block coding scheme.
A first observation is that,
without any loss of generality, we can restrict ourselves to the
case when the Burnashev coefficient $D$ is finite, since otherwise
the claim (\ref{maintheo1}) is trivially true.
The main result of this section is contained in Theorem \ref{invtheo} whose
proof will pass through a series of intermediate steps,
contained in Lemmas \ref{lemmaPMAP}, \ref{lemmaC}, \ref{lemmaD0} and \ref{lemmaC1}.

The main idea, taken from Burnashev's original paper
\cite{Burnashev} (see also \cite{GallagerNakiboglu} and
\cite{TelatarRimoldi}) is to obtain two different upper bounds for
the error probability.
Differently from \cite{Burnashev} and
\cite{GallagerNakiboglu}, we will follow an approach similar to
the one proposed in \cite{TelatarRimoldi} and look at the
behaviour of the a posteriori error probability, instead of that
of the a posteriori entropy.
The two bounds correspond to two distinct
phases which can be recognized in any sequential transmission
scheme and will be the content of Sections \ref{sect4.1} and
\ref{sect4.2}. The first one is provided in Lemma \ref{lemmaC}
whose proof is based on an application of Fano's inequality
combined with a martingale argument invoking Doob's optional
stopping theorem. The second bound is given by Theorem
\ref{lemmaC1} whose proof combines the use of the log-sum
inequality with another application of Doob's optional stopping
theorem. In Section \ref{sect4.3} these two
bounds will be combined obtaining Theorem \ref{invtheo} which is a
generalization to our setting of Burnashev's result
\cite{Burnashev}.

\subsection{A first bound on the error probability}
\label{sect4.1}

Suppose we are given a causal feedback encoder $\Phi=(\mc
W,(\phi_t))$ as in (\ref{causalencdef}) and a transmission time
$T$ as in Def. \ref{defdectime}. Our goal is to find a lower bound
for the error probability $p_e(\Phi,T,\Psi)$ where $\Psi$ is an
arbitrary sequential decoder for $\Phi$ and $T$.

It will be convenient to define for every $t\ge 0$ the
$\sigma$-algebra $\mc G_t:=\mc E_{t+1}$ describing the encoder's
feedback information at time $t+1$. $\mc G:=(\mc G_t)_{t\in\Z_+}$
will denote the corresponding filtration. We define the maximum a
posteriori error probability conditioned on the $\sigma$-algebras
$\mc F_t$ and $\mc G_t$ respectively by
$$P_{MAP}^{\Phi}(t):=1-\max_{w\in\mc W}\left\{\P_{\Phi}\left(W=w|\mc F_{t}\right)\right\}\,,
\qquad
\tilde P_{MAP}^{\Phi}(t):=1-\max_{w\in\mc W}\left\{\P_{\Phi}\left(W=w|\mc G_{t}\right)\right\}\,.$$
Clearly $P_{\Phi}(t)$ is an $\mc F_t$-measurable random variable
while $\tilde P_{\Phi}(t)$ is $\mc G_t$-measurable.

It is a well known fact that the decoder minimizing the error
probability over the class of fixed-length decoders
$\left\{\Psi:\mc S^t\times\mc Y^t\ra\mc W\right\}$ is the maximum
a posteriori one, defined by
$$\Psi^t_{MAP}(\mb s,\mb y)=
\argmax\limits_{w\in\mc W}\left\{\P_{\Phi} (W=w|\mb S_1^t=\mb
s,\mb Y_1^t=\mb y)\right\}\,, \qquad \mb s\in\mc S^{t},\,\mb
y\in\mc Y^t\,,$$ (with the convention for the operator $\argmax$
to arbitrarily assign one of the optimizing values in case of
non-uniqueness). It will be convenient to consider the larger
class of decoders $\left\{\Psi:\mc S^{t+1}\times\mc Y^t\ra\mc
W\right\}$ (differing from the previous one because of the
possible dependence on the state at time $t+1$); over such a
class, the optimal decoder is given by
$$\tilde\Psi^t_{MAP}(\mb s,\mb y)=
\argmax\limits_{w\in\mc W}\left\{\P_{\Phi}(W=w|\mb S_1^{t+1}=\mb
s,\mb Y_1^t=\mb y)\right\}\,, \qquad \mb s\in\mc S^{t+1},\,\mb
y\in\mc Y^t\,.
$$
It follows that for any $\Psi:\mc S^t\times\mc Y^t\ra\mc W$ we
have
$$p_e(\Phi,t,\Psi)\ge p_e(\Phi,t,\Psi^t_{MAP})\ge p_e(\Phi,t,\tilde\Psi^t_{MAP})=
\E_{\Phi}\left[\tilde P_{MAP}^{\Phi}(t)\right]\,.$$

The discussion above naturally generalizes from the fixed length
setting to the sequential one. In particular, given a stopping
time $T$ for the decoder filtration $\mc F$, we observe that,
since $\mc F_t\subseteq\mc G_{t}$ for every $t\ge0$, $T$ is also
stopping time for the filtration $\mc G$ and $\mc
F_{T}\subseteq\mc G_{T}$. It follows that the error probability of
the scheme $(\Phi,T,\Psi)$, where $\Psi$ is an arbitrary $\mc
F_{T}$-measurable $\mc W$-valued r.v., is lower bounded by that
corresponding to the sequential improved MAP decoder
$\tilde\Psi^{T}_{MAP}$, defined by
$$\tilde\Psi^{T}_{MAP}:=\argmax\limits_{w\in\mc W}
\left\{\P_{\Phi}\left(W=w|\mc G_{T}\right)\right\}\,.$$
Therefore we can conclude that
\be p_e\left(\Phi,T,\Psi\right)\ge \E_{\Phi}\left[\tilde
P_{MAP}^{\Phi}(T)\right]\label{pe>peMAP}\,,\ee for any $\mc
F_{T}$-measurable $\mc W$-valued random variable $\Psi$.

In the sequel we will lower bound the righthand side of
(\ref{pe>peMAP}). In particular, since the random variable $W$ is
uniformly distributed over the message set $\mc W$, and since
$S_1$ is independent of $W$, we have that
$$\P_{\Phi}\left(W=w|\mc G_0\right)=\P_{\Phi}(W=w)=\frac1{|\mc W|}\,,\qquad w\in\mc W\,,$$
so that
$$\tilde P^{\Phi}_{MAP}(0)=\frac{|\mc W|-1}{|\mc W|}\,.$$
Moreover we have the following recursive lower bound for $\tilde
P_{MAP}(t)$ (see Proposition 2 in \cite{TelatarRimoldi} for a
similar result in the memoryless case).
\begin{lemma}
\label{lemmaPMAP} Given any causal feedback encoder $\Phi$, we
have, for every $t$ in $\N$,
$$\tilde P^{\Phi}_{MAP}(t)\ge \lambda\tilde P_{MAP}^{\Phi}(t-1)\qquad \P_{\Phi}-a.s.$$
\end{lemma}
\proof See Appendix \ref{AppendixA}. $\hfill\square$

For every $\delta$ in $(0,\tfrac12)$, we now consider the random
variable \be\tau_{\delta}:=
\min\left\{T,\inf\left\{n\in\N:\,\tilde
P_{MAP}(t)\le\delta\right\}\right\}\,. \label{taudeltadef} \ee It
is immediate to verify that $\tau_{\delta}$ is a stopping time for
the filtration $\mc G$. Moreover the event $\left\{\tilde
P_{MAP}(\tau_{\delta})>\delta\right\}$ implies the event
$\left\{\tau_{\delta}=T\right\}$, so that an application of Markov
inequality and (\ref{pe>peMAP}) give us
$$
\ba{rcl} \P_{\Phi}\left(\tilde
P_{MAP}(\tau_{\delta})>\delta\right)&=&
\P_{\Phi}\left(\left\{\tilde P_{MAP}(\tau_{\delta})>\delta\right\}
\cap\left\{\tau_{\delta}=T\right\}\right)\\[5pt]
&\le&
\P_{\Phi}\left(\tilde P_{MAP}(T)>\delta\right)\\[5pt]
&\le& \frac1{\delta}\E_{\Phi}\left[\tilde P_{MAP}(T)\right]
\\[5pt]
&\le& \frac1{\delta}p_e(\Phi,T,\Psi)\,.\ea
$$

We introduce the following notation for the a posteriori entropy
$$\Gamma_t:=-\summ_{w\in\mc W}\P_{\Phi}\left(W=w|\,\mc G_t\right)
\log\P_{\Phi}\left(W=w|\,\mc G_t\right)\,,\qquad t\in\Z_+\,. $$
Observe that, since $S_1$ is independent of the message $W$, then
$$\Gamma_0=\log|\mc W|\,,\qquad \P_{\Phi}-a.s.$$
It is easy to verify that, whenever $\tilde
P_{MAP}(\tau_{\delta})\le\delta$, we have
$$\Gamma_{\tau_{\delta}}\le\ent(\delta)+\delta\log|\mc W|\,.$$
Hence the expected value of $\Gamma_{\tau_{\delta}}$ can be
bounded from above as follows: \be\label{E[h]<=} \ba{rcl}
\E_{\Phi}\left[\Gamma_{\tau_{\delta}}\right] &=&
\E_{\Phi}\left[\Gamma_{\tau_{\delta}}\big|\, \tilde
P_{MAP}(\tau_{\delta})\le\delta\right]
\P_{\Phi}\left(\tilde P_{MAP}(\tau_{\delta})\le\delta\right)\\[8pt]
&&+\, \E_{\Phi}\left[\Gamma_{\tau_{\delta}}\big|\, \tilde
P_{MAP}(\tau_{\delta})>\delta\right] \P_{\Phi}\left(\tilde
P_{MAP}(\tau_{\delta})>\delta\right)
\\[8pt]
&\le& \left(\ent(\delta)+\delta\log|\mc W|\right)
\P_{\Phi}\left(\tilde P_{MAP}(\tau_{\delta})\le\delta\right)+
\P_{\Phi}\left(\tilde P_{MAP}(\tau_{\delta})>\delta\right)
\log|\mc W|
\\[8pt]
&\le&
\ent(\delta)+\left(\delta+\frac1{\delta}{p_e(\Phi,T,\Psi)}\right)\log|\mc
W|\,. \ea \ee

We now introduce, for every time $t$ in $\N$, a $\mc P(\mc
X)$-valued random variable $\mb\Upsilon_{\Phi,t}$ describing the
channel input distribution induced by the encoder $\Phi$ at time
$t$: \be \mb\Upsilon_{\Phi,t}(x):=\P\left(X_t=x|\mc E_{t}\right) =
\P\left(\phi_t(W,\mb S_1^t,\mb Y_1^{t-1})=x|\,\mb S_1^t,\mb
Y_1^{t-1}\right)\,,\qquad x\in\mc X\,, \label{Upsolontdef} \ee
Notice that $\mb\Upsilon_{\Phi,t}$ is $\mc E_{t}$-measurable, i.e.
equivalently it is a function of the pair $(\mb S_1^t,\mb
Y_1^{t-1})$. The subscript in $\mb\Upsilon_{\Phi,t}$ emphasizes
the fact that this quantity depends on the encoder $\Phi$, with no
restriction on it but to be causal.


The following result relates three relevant quantities
characterizing the performances of any causal encoder sequential
decoder pair: the cardinality of the message set $\mc W$, the
error probability of the encoder decoder pair, and the the mutual
information cost $c$ (\ref{cdef}) up to the stopping time $\tau_{\delta}$:
\be
C_{\delta}(\Phi,T):=
\E_{\Phi}\left[\summ_{t=1}^{\tau_{\delta}}c(S_t,\mb\Upsilon_{\Phi,t})\right]\,.
\label{Cdeltadef}\ee
\begin{lemma}\label{lemmaC}
For any variable-length block-coding scheme
$\left(\Phi,T,\Psi\right)$ and any $0<\delta<\frac12$, we have \be
C_{\delta}(\Phi,T)\ge
\left(1-\delta-\frac{p_e\left(\Phi,T,\Psi\right)}{\delta}\right)
\log|\mc W|-\ent(\delta)\,. \label{Cineq}\ee
\end{lemma}
\proof See Appendix \ref{AppendixA}. $\hfill\square$

\subsection{A lower bound to the error probability of a composite binary hypothesis test}
\label{sect4.2}
%

We now consider a particular binary hypothesis testing problem
which will arise while proving the main result. Suppose we are
given a causal feedback encoder $\Phi=\left(\mc
W,(\phi_t)\right)$. Consider a nontrivial binary partition of the
message set \be\mc W=\mc W_0\cup\mc W_1\,,\qquad \mc W_0\cap\mc
W_1=\emptyset\,,\qquad \mc W_0,\mc
W_1\ne\emptyset\,,\label{partition}\ee and a sequential binary
hypothesis test $\tilde\Psi=(T,\tilde\Psi)$ (where $T$ is stopping
time for the filtration $\mc G$, and $\tilde\Psi$ is an $\mc
G_T$-measurable $\{0,1\}$-valued random variable) between the
hypothesis \mbox{$\{W\in\mc W_0\}$} and the hypothesis
\mbox{$\{W\in\mc W_0\}$}. Following the standard statistical
terminology we call $\tilde\Psi$ a composite test since it must
decide between two classes of probability laws for the process
$(\mb S,\mb Y)$ rather than between two single laws.
For every $t$, we define the $\mc P(\mc X)$-valued random
variables $\mb\Upsilon^0_{\Phi,t}$ and $\mb\Upsilon^1_{\Phi,t}$ by
$$
\mb\Upsilon^i_{\Phi,t}(x)= \P_{\Phi}\left(X_t=x|\,W\in\mc
W_i,\,\mc E_{t}\right)\,,\qquad x\in\mc X,\ i=0,1\,.
$$
The r.v. $\mb\Upsilon^0_{\Phi,t}$ (respectively
$\mb\Upsilon^1_{\Phi,t}$) represents the channel input
distribution at time $t$ induced by the encoder $\Phi$ when
restricted to the message subset $\mc W_0$ (resp. $\mc W_1$).
Notice that
$$\mb\Upsilon_{\Phi,t}=\P_{\Phi}(W\in\mc W_0|\mc E_t)\mb\Upsilon^0_{\Phi,t}+
\P_{\Phi}(W\in\mc W_1|\mc E_t)\mb\Upsilon^1_{\Phi,t}\,.$$

Let $\tau$ be another stopping time for the filtration $\mc G$,
such that $\tau\le T$. Let us consider the conditional expectation
terms
$$L_i:=\E_{\Phi}\left[\log\frac {\P_{\Phi}
\left(\mb S_{\tau+2}^{T+1},\mb Y_{\tau+1}^{T} \big|\,W\in\mc
W_i,\mc G_{\tau}\right)} {\P_{\Phi}\left(\mb S_{\tau+2}^{T+1},\mb
Y_{\tau+1}^{T} \big|\,W\notin\mc W_i,\mc G_{\tau}\right)}\Bigg|\,
W\in\mc W_i,\mc G_{\tau}\right]\,,\qquad i=0,1\,.$$ Both $L_0$ and
$L_1$ are $\mc G_{\tau}$-measurable random variables. In
particular $L_0$ equals the Kullback-Leibler information
divergence between the $\mc G_{\tau}$-conditioned probability
distributions of the pair $\left(\mb S_{\tau+2}^{ T+1},\mb
Y_{\tau+1}^{T}\right)$ respectively given $\{W\in\mc W_0\}$ and
$\{W\in\mc W_1\}$; an analogous interpretation is possible, mutati
mutandis, for $L_1$.

In the special case when both $\tau$ and $T$ are deterministic
constants, an application of the log-sum inequality would show
that, for $i=0,1$, $L_i$ can be upperbounded by the $\mc
G_{\tau}$-conditional expected value of the sum of the information
divergence costs $d\left(\mb\Upsilon^i_{\Phi,t},S_t\right)$ from
time $\tau+1$ to $T$, and analogously for $L_1$, with the terms
$d\left(S_t,\mb\Upsilon^i_{\Phi,t}\right)$. It turns out that the
same is true in our setting where both $\mc\tau$ and $T$ are
stopping times for the filtration $\mc G$, as stated in the
following lemma, whose proof requires, besides an application of
the log-sum inequality, a martingale argument invoking Doob's
optional stopping theorem.

\begin{lemma}\label{lemmaD0}
Let $\tau$ and $T$ be stopping times for the filtration $\mc G$
such that $\tau\le T$, and consider a partition of the message set
as in (\ref{partition}). Then \be
L_i\le\E_{\Phi}\left[\summ_{t=\tau+1}^{T}
d\left(\mb\Upsilon_{\Phi,t}^i,S_t\right) \Big|\,W\in\mc W_i,\mc
G_{\tau}\right]\,,\qquad\P_{\Phi}-a.s.\,,\ i=0,1\,.
\label{Wi<=K}\ee
\end{lemma}
\proof See Appendix \ref{AppendixA}. $\hfill\square$

Suppose now that $\mc W_1$ is a $\mc G_{\tau}$-measurable random
variable taking values in $2^{\mc W}\setminus\{\emptyset,\mc W\}$,
the class of nontrivial proper subsets of the message set $\mc W$.
In other words, we are assuming that $\mc W_1$ is a random subset
of the message set $\mc W$, deterministic function of the pair
$(\mb S_1^{\tau+1},\mb Y_1^{\tau})$.

The following result gives a lower bound on the error probability
of the binary test $\tilde\Psi$ conditioned on the
$\sigma$-algebra $\mc G_{\tau}$ in terms of the information divergence
terms
$$\E_{\Phi}\left[\summ_{t=\tau+1}^Td\left(S_t,\mb\Upsilon^i_{\Phi,t}\right)\big|\,\mc G_{\tau}\right]
\,,\qquad i=0,1\,.$$

\begin{lemma}
\label{lemmaC1} Let $\Phi$ be any causal encoder, and $\tau$ and
$T$ be stopping times for the filtration $\mc G$ such that
$\tau\le T$. Then, for every $2^{\mc W}\setminus\{\emptyset,\mc W\}$-valued $\mc
G_{\tau}$-measurable r.v. $\mc W_1$, we have that $\P_{\Phi}-a.s.$
\be\label{proC1ineq}
\E_{\Phi}\left[\summ_{t=\tau+1}^Td\left(S_t,\mb\Upsilon^{\1_{\{W\in\mc W_1\}}}_{\Phi,t}\right)\big|\,\mc G_{\tau}\right]
\ge
\log\frac{Z}{4}-\log\P\left(\tilde\Psi\ne\1_{\{W\in\mc W_1\}}
\big|\,\mc G_{\tau}\right)\ee  where
$$Z:=\min\Big\{
\P_{\Phi}\left(W\in\mc W_0|\,\mc G_{\tau}\right)\,,\
\P_{\Phi}\left(W\in\mc W_1|\,\mc G_{\tau}\right) \Big\}\,.$$
\end{lemma}
\proof See Appendix \ref{AppendixA}. $\hfill\square$
\subsection{Burnashev bound for Markov channels}
\label{sect4.3}
\begin{lemma}\label{lemmaadded}
Let $\Phi$ be a causal feedback encoder and $T$ a transmission time for $\Phi$.
Then, for every $0<\delta<1/2$ there exists a
$\mc G_{\tau_{\delta}}$-measurable random subset $\mc W_1$ of the
message set $\mc W$, whose a posteriori error probabilities
satisfy \be1-\lambda\delta\ge\P\left(W\in\mc W_1\big|\mc
G_{\tau_{\delta}}\right)\ge\lambda\delta\,.\label{>=lambdadelta}\ee
\end{lemma}
\proof See Appendix \ref{AppendixA}.$\hfill\square$

To a causal encoder $\Phi$ and a transmission time $T$,
for every $0<\delta<1/2$ we define the quantity
\be\label{Ddeltadef}
D_{\delta}(\Phi,T):=\max_{\substack{\mc W_1\ \mc G_{\tau_{\delta}}-measurable\\
2^{\mc W}-valued\ r.v.\\
\lambda\delta\le\P_{\Phi}(W\in\mc W_1|\mc G_{\tau_{\delta}})\le1-\lambda\delta}}
\E\left[\summ_{t=\tau_{\delta}+1}^Td\left(S_t,\mb\Upsilon^{\1_{\{W\in\mc W_1\}}}_{\Phi,t}\right)\right]
\ee
The quantity $D_{\delta}(\Phi,T)$ equals the maximum,
over all possible choices of a nontrivial partition of the message set $\mc W$
as a function of the joint channel state output process
$(\mb S_1^{\tau_{\delta+1}},\mb Y_1^{\tau_{\delta}})$ stopped
at the intermediate time $\tau_{\delta}$, of the averaged
sum of the information divergence costs $d\left(S_t,\mb\Upsilon^{\1_{\{W\in\mc W_1\}}}\right)$
incurred between times $\tau_{\delta}+1$ and $T$. Intuitively
$D_{\delta}(\Phi,T)$ measures the maximum error exponent achievable
by the encoder $\Phi$ when transmitting a binary message between times $\tau_{\delta}$
and $T$.

Based on Lemma \ref{lemmaC} and Lemma
\ref{lemmaC1}, we will now prove the main result of this section,
consisting in an upper bound on the largest error exponent achievable
by variable-length block-coding schemes
with perfect causal state knowledge and output feedback.

\begin{theorem}
\label{invtheo} Consider a variable-length block-coding scheme
$\left(\Phi,T,\Psi\right)$. Then, for every
$\delta\in(0,\tfrac12)$,
\be \ds-\log p_e\left(\Phi,T,\Psi\right) \le \ds
\frac{D}{C}C_{\delta}(\Phi,T)+ D_{\delta}(\Phi,T) - \frac{D}{C}\log|\mc
W|\left(1-\alpha\right)-\beta\,, \label{invtheoclaim} \ee where
$$
\alpha:=\delta+\frac{p_e(\Phi,T,\Psi)}{\delta} \,,\qquad
\beta:=\log\frac{\lambda\delta}4-\frac{D}{C}\ent(\delta)\,.$$
\end{theorem}
\proof
Let $\mc W_1$ be a $\mc G_{\tau_{\delta}}$-measurable subset of the message space $\mc W$
satisfying (\ref{>=lambdadelta}).
We define the binary sequential decoder
$$\tilde\Psi_{\delta}:=\1_{\mc W_1}\left(\Psi\right)\,.$$
Notice that the definition above is consistent in the sense that
$\tilde\Psi$ is $\mc G_{T}$-measurable, since $\Psi$ is $\mc
G_{T}$-measurable, while $\mc W_1$ is $\mc
G_{\tau_{\delta}}$-measurable and $\mc
G_{\tau_{\delta}}\subseteq\mc G_{T}$.

We can lower bound the error probability of the composite
hypothesis test $\tilde\Psi_{\delta}$ conditioned on $\mc
G_{\tau_{\delta}}$ using Lemma \ref{lemmaC1} and
(\ref{>=lambdadelta}), obtaining
$$
\ba{rcl} -\log\P_{\Phi}\left(\tilde\Psi_{\delta}\ne \1_{\mc
W_1}(W)\,\big|\,\mc G_{\tau_{\delta}}\right)+
\log\frac{\lambda\delta}4 &\le&
\E_{\Phi}\left[\summ_{t=\tau_{\delta}+1}^{T}
d\left(S_{t},\mb\Upsilon_{\Phi,t}^{\1_{\{W\in\mc W_1\}}}\right)
\big|\,\mc G_{\tau}\right]\,.\ea
$$
It is clear that the error event of the pair $\Psi$ is implied by
the error event of $\tilde\Psi_{\delta}$. It follows that \be
\ba{rcl} -\log p_e\left(\Phi,T,\Psi\right)
+\log\ds\frac{\lambda\delta}4&=& -\log
\E_{\Phi}\left[\P\left(\Psi\ne W\big|\,\mc
G_{\tau_{\delta}}\right)\right]
+\log\ds\frac{\lambda\delta}4\\[8pt]
&\le&\ds -\log\E_{\Phi}\left[ \P\left(\1_{\mc W_1}\left(\Psi\right)\ne
\1_{\mc W_1}(W)\big|\,\mc G_{\tau_{\delta}}\right) \right]
+\log\frac{\lambda\delta}4\\[8pt]
&=&\ds -\log\E_{\Phi}\left[ \P\left(\tilde\Psi_{\delta}\ne
\1_{\mc W_1}(W)\big|\,\mc G_{\tau_{\delta}}\right) \right]
+\log\frac{\lambda\delta}4\\[8pt]
&\le&\ds \E_{\Phi}\left[-\log \P_{\Phi}
\left(\tilde\Psi_{\delta}\ne \1_{\mc W_1}(W)\big|\,\mc
G_{\tau_{\delta}}\right)\right]
+\log\frac{\lambda\delta}4\\[8pt]
&\le&\E_{\Phi} \left[
\E_{\Phi}\left[\summ_{t=\tau_{\delta}+1}^Td\left(S_t,\mb\Upsilon^{\1_{\{W\in\mc W_1\}}}_{\Phi,t}\right)|\,
\mc G_{\tau_{\delta}}\right]
\right]
\\[13pt]&\le&
D_{\delta}(\Phi,T)
\,,
\ea\label{invtheoII}
\ee
the second inequality in (\ref{invtheoII}) following from Chebychev inequality. The claim
now follows by taking a linear combination of (\ref{invtheoII})
and (\ref{Cineq}). \qed

In the memoryless case, i.e. when the state space is trivial
($|\mc S|=1$), Burnashev's original result (see (4.1) in
\cite{Burnashev}, see also (12) in \cite{TelatarRimoldi}) can be
recovered from $(\ref{invtheoclaim})$ by optimizing over the
channel input distributions $\mb\Upsilon_{\Phi,t}$,
$\mb\Upsilon^0_{\Phi,t}$, and $\mb\Upsilon^1_{\Phi,t}$.

In order to prove Part 1 of Theorem \ref{maintheo} it remains to
consider countable families of variable-length coding schemes with
vanishing error probability and to show that asymptotically the
upper bound in (\ref{invtheoclaim}) reduces to the Burnashev
exponent $E_B(R)$. This involves new technical challenges which
will be the object of next section.


\section{Markov decision problems with stopping time horizons}
\label{sect3}

In this section we shall recall some concepts about Markov
decision processes which will allow us to asymptotically estimate
the terms $C_{\delta}(\Phi,T)$ and $D_{\delta}(\Phi,T)$
respectively in terms of the capacity $C$
defined in (\ref{capacitydef}) and the Burnashev coefficient $D$
(\ref{Burnashevcoeffdef}) of the FSMC.

The main idea is to interpret the maximization of
$C_{\delta}(\Phi,T)$ and $D_{\delta}(\Phi,T)$ as
stochastic control problems with average cost criterion
\cite{Aristotele}. The control is the channel input distribution
chosen as a function of the available feedback information and the
controller is identified with the encoder. The main novelty these
problems have with respect to those traditionally addressed by
Markov decision theory consists in the fact that, as a consequence
of considering variable-length coding schemes, we shall need to
deal with the situation when the horizon is neither finite (in the
sense of being a deterministic constant) nor infinite (in the
sense of being concerned with the asymptotic normalized average
running cost), but rather it is allowed to be a random stopping
time. In order to handle this case we adopt the convex analytical
approach, a technique first introduced by Manne in \cite{Manne}
(see also \cite{Derman}) for the finite state finite action
setting, and later developed in great generality by Borkar
\cite{BorkarConvexAnalytical}.

In Section \ref{sect3.1} we shall first reformulate the problem of
optimizing the terms $C_{\delta}(\Phi,T)$ and $D_{\delta}(\Phi,T)$
with respect to the causal encoder $\Phi$.
Then, we present a brief review of the convex
analytical approach to Markov decision problems in Section
\ref{sect3.2}, presenting the main ideas and definitions. In
Section \ref{sect3.3} we will prove a uniform convergence theorem
for the empirical measure process and use this result to treat the
asymptotic case of the average cost problem with stopping time
horizon. The main result of this section is contained in Theorem
\ref{lemmac*}, which is then applied in Section \ref{sect4.4} together with
Theorem \ref{invtheo} in order to prove Part 1 of Theorem \ref{maintheo}.

\subsection{Markov decision problems with stopping time horizons}
\label{sect3.1} We shall consider a controlled Markov chain over
$\mc S$, with compact control space $\mc U:=\mc P(\mc X)$, the
space of channel input distributions. Let $g:\mc S\times\mc
U\ra\R$ be a continuous (and thus bounded) cost function; in our
application $g$ will coincide either with the mutual information
cost $c$ defined in (\ref{cdef}) or with the information
divergence cost $d$ defined in (\ref{ddef}). We prefer to consider
the general case in order to deal with both problems at once.

The evolution of the system is described by a state sequence $\mb
S=(S_t)$, an output sequence $\mb Y=(Y_t)$ and a control sequence
$\mb U=(U_t)$. If at time $t$ the system is in state $S_t=s$ in
$\mc S$, and a control $U_t=u$ in $\mc U$ is chosen according to
some policy, then a cost $g(s,u)$ is incurred and the system
produces the output $Y_t=y$ in $\mc Y$ and moves to next state
$S_{t+1}=s_+$ in $\mc S$ according to the stochastic kernel
$Q(s_+,y|\,s,u)$, defined in (\ref{Qdef}).
Once the transition into next state has occurred, a
new action is chosen and the process is repeated.

At time $t$, the control $U_t$ is allowed to be an $\mc
E_t$-measurable random variable, where $\mc E_t=\sigma(\mb
S_1^t,\mb Y_1^{t-1})$ is the encoder's feedback information
pattern at time $t$; in other words we are assuming that
$U_t=\pi_t\left(\mb S_1^t,\mb Y_1^{t-1}\right)$ for some map
$$\pi_t:\mc S^t\times\mc Y^{t-1}\ra\mc U\,.$$ We define a \emph{feasible policy} $\mb\pi$
as a sequence $(\pi_t)_{t\in\N}$ of such maps. Once a feasible
policy $\mb\pi$ has been chosen, a joint probability distribution
$\P_{\mb\pi}$ for state, control and output sequences is well
defined; we will denote by $\E_{\mb\pi}$ the corresponding
expectation operator.


Let $\tau$ be a stopping time for the filtration $\mc G=(\mc G_t)$
(recall that \mbox{$\mc G_t=\mc E_{t+1}$} describes the encoder's
feedback and state information at time $t+1$),
and consider the following optimization problem: maximize
\be\frac1{\E_{\mb\pi}[\tau]}\E_{\mb\pi}\left[\summ_{t=1}^{\tau}
g\left(S_t,\pi_t(\mb S_1^t,\mb Y_1^{t-1})\right)\right]
\label{stoppingoptmal}\ee over all feasible policies
$\mb\pi=(\pi_t)$ such that $\E_{\mb\pi}[\tau]$ is finite.

Clearly, in the special case when $\tau$ is a constant
(\ref{stoppingoptmal}) reduces to the standard finite-horizon
problem which is usually solved with dynamic programming tools.
Another special case is when $\tau$ is geometrically distributed and
independent from the processes $\mb S$, $\mb U$ and $\mb Y$. In
this case (\ref{stoppingoptmal}) reduces to the so-called
discounted problem which has been widely studied in the stochastic
control literature \cite{Aristotele}. However, what makes the
problem nonstandard is that in (\ref{stoppingoptmal}) $\tau$ is
allowed to be an arbitrary stopping time for the filtration $\mc
F$, generally correlated with the processes $\mb S$, $\mb U$ and
$\mb Y$.


\subsection{The convex analytical approach}
\label{sect3.2} We review some of the ideas of the
convex-analytical approach following
\cite{BorkarConvexAnalytical}.

A feasible policy $\mb\pi$ is said to be stationary if the current
control depends on the current state only and is independent of
the past state and output history and of the time, i.e. there
exists a map $\pi:\mc S\ra\mc U$ such that $\pi_t(\mb s_1^t,\mb
y_1^{t-1})=\pi(s_t)$ for all $t$. We will identify a stationary
policy as above with the map $\pi:\mc S\ra\mc U$. It has already
been noted in Section \ref{sect2.1} that, for every stationary
policy $\pi$, the stochastic matrix $Q_{\pi}$ as defined in
(\ref{Qpidef}) is irreducible, so that existence and uniqueness of an
invariant measure $\mb\mu_{\pi}$ in $\mc P(\mc S)$ are guaranteed.
It follows that, if a stationary
policy $\pi$ is used, then the normalized running cost
$\frac1n\summ_{t=1}^ng(S_t,\pi(S_t))$ converges $\P_{\pi}$-almost surely to
$\summ_{s\in\mc S}\mb\mu_{\pi}(s)g(s,\pi(s))$. Define \be
G:=\max_{\pi:\mc S\ra\mc U}\summ_{s\in\mc
S}\mb\mu_{\pi}(s)g(s,\pi(s))\,. \label{Gdef}\ee Observe that the
optimization in the righthand side of (\ref{Gdef}) has the same
form of those in the definitions (\ref{cdef}) and
(\ref{Burnashevcoeffdef}) of the capacity and the Burnashev
coefficient of an ergodic FSMC given in Section \ref{sect2}.
Notice that compactness of the space $\mc U^{\mc S}$ of all
stationary policies and continuity of the cost $g(s,\pi(s))$ and
of the invariant measure $\mb\mu_{\pi}$ as functions of the
stationary policy $\pi$ guarantee the existence of an optimal
value in the above maximization.

We now consider stationary randomized policies. These are defined
as maps $\tilde{\pi}:\mc S\ra\mc P(\mc U)$, where $\mc P(\mc U)$
denotes the space of probability measures on $\mc U$, equipped
with its Prohorov topology \cite{Borkarbook}. To any stationary
randomized policy $\tilde{\pi}$ the following control strategy is
associated: if at time $t$ the state is $S_t=s$, then the control
$U_t$ is randomly chosen in the control space $\mc U$ with
conditional distribution given the available information
$\mc E_t=\sigma(\mb S_1^t,\mb Y_1^{t-1})$ equal to $\pi(s)$.
Observe that there are two levels of randomization.
The control space itself has already been defined
as the space of channel input probability distributions $\mc P(\mc
X)$, while the strategy associated to the stationary randomized
policy $\tilde{\pi}$ chooses a control at random with conditional
distribution $\tilde{\pi}(S_t)$ in $\mc P(\mc U)=\mc P(\mc P(\mc
X))$. Of course randomized stationary policies are a
generalization of deterministic stationary policies, since to any
deterministic stationary policy $\pi:\mc S\rightarrow\mc U$ it is
possible to associate the randomized policy
$\tilde{\pi}(s)=\delta_{\pi(s)}$. To any randomized stationary
policy $\tilde{\pi}:\mc S\ra\mc P(\mc U)$ we associate the
stochastic matrix describing the associated state transition
probabilities
\be\left(Q_{\tilde{\pi}}(s_+|\,s)\right)_{s,s_+\in\mc S}\,,\qquad
Q_{\tilde{\pi}}(s_+|\,s):=\int_{\mc
U}Q(s_+|\,s,u)[\tilde{\pi}](s)\left(\de
u\right)\,.\label{Qpidef2}\ee Similarly to the case of stationary
deterministic policies, it is not difficult to conclude that,
since $Q_{\tilde{\pi}}$ can be written as a convex combination of
a finite number of stochastic matrices $Q_{f}$, with $f:\mc
S\ra\mc X$, all of which are irreducible, then $Q_{\tilde{\pi}}$
itself is irreducible and thus admits a unique state ergodic
measure $\mb\mu_{\tilde{\pi}}$ in $\mc P(\mc S)$. This motivates
the following definition.

\begin{figure}
\centering
\includegraphics[height=4.0cm,width=4.0cm]{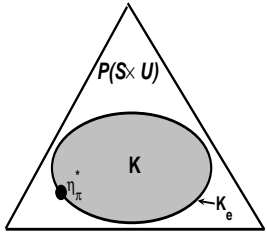}
\caption{A schematic representation of the optimization problem
(\ref{convexoptproblem}). The large triangular space is the
infinite dimensional Prohorov space $\mc P(\mc S\times\mc U)$. Its
gray-shaped subset represents the close convex set $K$ of all
occupation measures. The set of extreme points of $K$ is $K_e$ and
corresponds to the set of all occupation measures associated to
stationary deterministic policies. The optimal value of the linear
functional $\mb\eta\mapsto\langle\mb\eta,g\rangle$ happens to be
achieved on $K_e$ and thus corresponds to the occupation measure
$\mb\eta_{\pi}^*$ associated to an optimal deterministic
stationary policy $\pi^*:\mc S\rightarrow\mc P(\mc X)$. }
\label{figureconvexanalytical}
\end{figure}

\begin{definition}
For every stationary (randomized) policy $\pi:\mc S\ra\mc P(\mc
U)$ the {\emph occupation measure} of $\pi$ is $\mb\eta_{\pi}$ in
$\mc P(\mc S\times\mc U)$ defined by
$$\langle \mb\eta_{\pi},h\rangle=
\int_{\mc S\times\mc U} h(s,u)\de{\mb\eta}_{\pi}(s,u)=
\summ_{s\in\mc S}\mb\mu_{\pi}(s)\int_{\mc U}h(s,u)
\de\pi(s)\,,\qquad \forall\,h\in\,\mc C_b(\mc S\times\mc U)\,,$$
where $\mb\mu_{\pi}$ in $\mc P(\mc S)$ is the invariant measure of
the stochastic matrix $Q_{\pi}$, while $\mc C_b(\mc S\times\mc U)$
is the space of bounded continuous maps from $\mc S\times\mc U$ to
$\R$.
\end{definition}
The occupation measure $\mb\eta_{\pi}$ can be viewed as the
long-time empirical frequency of the joint state-control process
governed by the stationary (randomized) policy $\pi$. In fact, for
every time $n$ in $\N$, we can associate to the controlled Markov
process the empirical measure $\mb\ups_n$ which is a $\mc P(\mc
S\times\mc U)$-valued random variable sample-path-wise defined by
\be\langle\mb\ups_n,h\rangle:=\frac1n\summ_{t=1}^nh(S_t,U_t)\,,\qquad
\forall\ h\in\mc C_b(\mc S\times\mc U)\,.\label{ups_ndef}\ee
Observe that $\mb\ups_n$ is a probability measure on the product
space $\mc S\times\mc U$, and is itself a random variable since
it is defined as a function of the joint state control random process
$(\mb S_1^t,\mb U_1^t)$.
Then, it can be verified that, if the process is controlled by a
stationary (randomized) policy $\pi$, then \be
\lim_{n\in\N}\mb\ups_n=\mb\eta_{\pi}\,\qquad
\P_{\pi}-a.s.\label{ups_n->eta_pi}\ee

We will denote by $K$ the set of the occupation measures
associated to all the stationary randomized policies, i.e. \be
K:=\left\{{\mb\eta}_{\tilde{\pi}}\,|\,\tilde{\pi}:\mc S\ra\mc
P(\mc U)\right\} \subseteq\mc P(\mc S\times\mc U), \label{Kdef}\ee
and by $K_e$ the set of all occupation measures associated to
stationary deterministic policies
$$K_e:=\left\{{\mb\eta}_{\pi}\,|\,\pi:\mc S\ra\mc U\right\}\subseteq\mc P(\mc S\times\mc U)\,.$$
Well known results (see \cite{BorkarConvexAnalytical}) show that
both $K$ and $K_e$ are closed subsets of $\mc P(\mc S\times\mc
U)$. Moreover $K$ is convex and $K_e$ coincides with the set of
extreme points of $K$. Furthermore it is possible to characterize
$K$ as the the set of zeros of the continuous linear functional
$$\mb F:\mc P(\mc S\times\mc U)\ra[0,1]^{\mc S}\,,\qquad
F_s(\mb\eta):=\mb\eta(\{s\},\mc U)-\int\limits_{\mc S\times\mc U}
Q_{S}(s\,|\,j,u)\de\mb\eta(j,u)\,,
$$
i.e. \be K=\left\{\mb\eta\in\mc P(\mc S\times\mc U): \mb
F(\mb\eta)=\mb 0\right\}\,. \label{Kcharacterization}\ee In fact
it is possible to think of $||\mb F(\mb\eta)||$ (here and
throughout the paper $||\mb x||:=\max_i |x_i|$ will denote the
$L^{\infty}$-norm of a vector $\mb x$) as a measure of how far the
$\mc S$-marginal of a measure $\mb\eta$ in $\mc P(\mc S\times\mc
U)$ is from being invariant for the state process.

If one were interested in optimizing the infinite-horizon running
average cost
$$\liminf_{n\in\N}\frac1n\E_{\pi}\left[\summ_{t=1}^ng(S_t,U_t)\right]=
\liminf_{n\in\N}\E_{\pi}\left[\langle\mb\ups_n,g\rangle\right]$$
over all (randomized) stationary policies $\pi$, then
(\ref{ups_n->eta_pi}) and (\ref{Kdef}) would immediately lead to
the following convex optimization problem: \be
\max\limits_{\substack{\vspace{2pt}\\ \mb\eta\in
K}}\langle\mb\eta,g\rangle\,.\label{convexoptproblem}\ee In fact,
using (\ref{Kcharacterization}), (\ref{convexoptproblem}) can be
rewritten as an infinite dimensional linear programming problem
\be \max_{\substack{\vspace{2pt} \\ \vspace{2pt}\mb\eta\in\mc
P(\mc S\times\mc U):\\\mb F(\mb\eta)=\mb0}}
\langle\mb\eta,c\rangle \,.\label{linearprogramming}\ee We notice
that, since $\mc U$ is compact and $\mc S$ is finite, $\mc P(\mc
S\times\mc U)$ is compact. Thus both $K$ and $K_e$ are compact. It
follows that, since the map
$$\mc P(\mc S\times\mc U)\ni\mb\eta\longmapsto\langle \mb\eta,g\rangle\in\R$$
is continuous, it achieves its maxima both on $K$ and $K_e$;
moreover, the same map is linear so that these maxima do coincide,
i.e. the maximum over $K$ is achieved in an extreme point. Thus we
have the following chain of equalities \be \ba{rcl} G&=&
\max\limits_{\substack{\vspace{2pt}\\ \pi:\mc S\rightarrow\mc U}}
\summ_{s\in\mc S}\mb\mu_{\pi}(s)g(s,\pi(s))\\[10pt]
& =& \ds
\max\limits_{\substack{\vspace{2pt}\\ \pi:\mc S\rightarrow\mc
U}}\langle\mb\eta_{\pi},g\rangle\\[5pt] &=& \ds
\max\limits_{\substack{\vspace{2pt}\\ \mb\eta\in
K_e}}\langle\mb\eta,g\rangle \\[5pt] &=& \ds
\max\limits_{\substack{\vspace{2pt}\\ \mb\eta\in
K}}\langle\mb\eta,g\rangle \\[5pt] &=& \ds\max_{\substack{\vspace{2pt} \\
\vspace{2pt}\mb\eta\in\mc P(\mc S\times\mc U):\\\mb
F(\mb\eta)=\mb0}} \langle\mb\eta,c\rangle\,.\ea\label{chain}\ee
We observe that the last term in (\ref{chain}) both the constraints and the object
functionals are linear.
This indicates (infinite dimensional) linear programming
as a possible approach for computing $G$, alternative to the dynamic programming
ones based on policy or value iteration techniques \cite{Aristotele},
\cite{BorkarConvexAnalytical}. Moreover, it shows an easy way to generalize the
theory taking into account average cost constraints (see \cite{GallagerNakiboglu}
where the Burnashev exponent of DMCs with average cost constraints is studied).
In fact, in the convex analytical approach these constraints merely translate into
additional constraints for the linear program.

\subsection{An asymptotic solution to Markov decision problems with a stopping time horizon}
\label{sect3.3}

It is known that, under the ergodicity and continuity assumptions
we have made, $G$ defined in (\ref{Gdef}) is the sample-path
optimal value for the infinite horizon problem with cost $g$ not
only over the set of all stationary policies, but also over the
larger set of all feasible policies (actually over all admissible
policies, see \cite{BorkarConvexAnalytical}). This means that, for
every feasible policy $\mb\pi=(\pi_t)$,
\be\limsup\limits_{n\in\N}\frac1n\summ_{t=1}^ng\left(S_t,\pi_t(\mb
S_1^t,\mb Y_1^{t-1})\right) \le G\,,\qquad \P_{\mb\pi}-a.s.\,.
\label{optimality1}\ee Moreover, it is a known fact that for an
arbitrary sequence of policies $(\mb\pi^n)$ we have
\be\limsup\limits_{n\in\N}\frac1n\E_{\mb\pi^n}\left[\summ_{t=1}^ng(S_t,U_t)\right]=
\limsup\limits_{n\in\N}\frac1n\E_{\mb\pi^n}\left[\summ_{t=1}^n
g\left(S_t,\pi^n_t\left(\mb S_1^t,\mb Y_1^{t-1}\right)\right)\right]
\le G\,, \label{limsupEn<=C}\ee i.e. the limit of the optimal
values of finite horizon problems coincides with infinite horizon
optimal value. (\ref{limsupEn<=C}) can be proved by using dynamic
programming arguments based on Bellman principle of optimality. As
shown in \cite{TatikondaMitter}, (\ref{limsupEn<=C}) is useful in
characterizing the capacity of channels with memory and feedback
with fixed-length codes. Actually, a much more general result than
(\ref{limsupEn<=C}) can be proved, as explained in the sequel.

In the convex analytical approach, the key point in proof of (\ref{optimality1})
consists in showing that, under any -generally
non stationary- feasible policy $\mb\pi$, the empirical measure
process $(\mb\ups_n)$ as defined in (\ref{ups_ndef}) converges
$\P_{\mb\pi}$-almost surely to the set $K$. The way this is
usually proven is by using a martingale central limit theorem in
order to show that the finite-dimensional process $\mb
F(\mb\ups_n)$ converges to $\mb 0$ almost surely. The following is
a stronger result, providing an exponential upper bound on the
tails of the random sequence $(||\mb F(\mb\ups_n)||)_n$, this
bound being uniform with respect to the choice of the policy
$\mb\pi$ in $\Pi$.
\begin{lemma}
\label{lemmaAzuma1} For every $\eps>0$, and for every feasible
policy $\mb\pi$ \be\P_{\mb\pi}\left(||\mb
F(\mb\ups_n)||\ge\eps+\frac1n\right)\le 2|\mc
S|\exp\left(-n\eps^2/2\right)\,. \label{Azuma1}\ee
\end{lemma}
\proof See Appendix \ref{AppendixB}.$\hfill\square$

We emphasize the fact that the bound (\ref{Azuma1}) is uniform
with respect to the choice of the feasible policy $\mb\pi$. It is
now possible to drive conclusions on the tails of the running
average cost $\frac1n\sum_{t=1}^ng(S_t,U_t)$ based on
(\ref{Azuma1}). The core idea is the following. By the definition
of the empirical measure $\mb\ups_n$, we can rewrite the
normalized running cost as
$$\frac1n\summ_{t=1}^ng(S_t,U_t)=\langle\mb\ups_n,g\rangle\,.$$
Since the map $\mb\eta\mapsto\langle\mb\eta,g\rangle$ is
continuous over $\mc P(\mc S\times\mc U)$, and
$G=\max\{\langle\mb\eta,g\rangle|\,\mb\eta\in K\}$, we have that,
whenever $\mb\nu_n$ is close to the set $K$,
$\langle\mb\nu_n,g\rangle$ cannot be much larger than $G$. It
follows that, if with high probability $\mb\ups_n$ is close enough
to $K$, then with high probability $\langle\mb\ups_n,g\rangle$
cannot be much larger than $G$. In order to show that with high
probability $\mb\ups_n$ is close to $K$, we want to use
(\ref{Azuma1}). In fact, if for some $x$ in $\mc P(\mc S\times\mc
U)$ the quantity $||\mb F(x)||$ is very small, then $x$ is
necessarily close to $G$. More precisely, we define the function
$$\gamma:\R^+\ra\R\,,\qquad
\gamma(x):=\sup\left\{\langle\mb\eta,g\rangle\big|\,\mb\eta\in\mc
P(\mc S\times\mc U):\, ||\mb F(\mb\eta)||\le x\right\}\,.$$
Clearly $\gamma$ is nondecreasing and $\gamma(0)=G$. Moreover we
have the following result.
\begin{lemma}
\label{lemmasemicontinuity} The map $\gamma$ is upper
semicontinuous. (i.e. $x_n\ra
x\Rightarrow\limsup_n\gamma(x_n)\le\gamma(x)$)
\end{lemma}
\proof See Appendix \ref{AppendixB}.\qed

For every $k$ in $\N$ we now introduce the random process
$(G^k_n)$
$$ G_n^k:=\sup\limits_{t\ge n}\langle\mb\ups_n,g\rangle\,,\qquad n\in\N\,.$$
Clearly the process $(G_n^k)$ is samplepathwise non increasing in
$n$.

\begin{lemma}\label{lemmanew}
Let $(\tau_k)$ be a sequence of stopping times for the filtration
$\mc F$ and $(\mb\pi^k)$ be a sequence of feasible policies such
that $\E_{\mb\pi^k}[\tau_k]<\infty$ for every $k$ and
\be \lim_{k\in\N}\P_{\mb\pi^k}\left(\tau_k\le
M\right)=0\,,\qquad \forall\, M\in\N\,.
\label{probabilisticdivergence}\ee Then \be
\lim_{k\in\N}\P_{\mb\pi^k}\left(G^k_{\tau_k}>\gamma(\eps)\right)=0\,,\qquad
\forall\eps>0\,. \label{PGtauk->0}\ee
\end{lemma}
\proof See Appendix \ref{AppendixB}.$\hfill\square$

The following result can be considered as an asymptotic estimate
of (\ref{stoppingoptmal}). It consists in a generalization of
(\ref{limsupEn<=C}) from a deterministic increasing sequence of
time horizons to a sequence of stopping times satisfying the
'probabilistic divergence' requirement
(\ref{probabilisticdivergence}).
\begin{theorem}\label{lemmac*}
Let $(\tau_k)$ be a sequence of stopping times for the filtration
$\mc F$ and $(\mb\pi^k)$ be a sequence of feasible policies  such
that $\E_{\mb\pi^k}[\tau_k]<\infty$ for every $k$, and (\ref{probabilisticdivergence}) holds true.
 Then \be \label{lemmac*claim}
\limsup_{k\in\N}\frac1{\E_{\mb\pi^k}\left[\tau_k\right]}
\E_{\mb\pi^k}\left[\summ_{t=1}^{\tau_k}g(S_t,U_t)\right]\le G\,.
\ee
\end{theorem}
\proof Let us fix an arbitrary $\eps>0$. By applying Lemma
\ref{lemmanew}, we obtain
$$
\ba{c}\ds \frac
{\E_{\mb\pi^k}\left[\summ_{t=1}^{\tau_k}g(S_t,U_t)\right]}
{\E_{\mb\pi^k}\left[\tau_k\right]} =\ds \frac
{\E_{\mb\pi^k}\left[\tau_k\langle\mb\ups_{\tau_k},g\rangle\right]}
{\E_{\mb\pi^k}\left[\tau_k\right]}\\[15pt]
=\ds \frac
{\E_{\mb\pi^k}\left[\tau_k\langle\mb\ups_{\tau_k},g\rangle
\1_{\{G^k_{\tau_k}\le\gamma(\eps)\}}\right]}
{\E_{\mb\pi^k}\left[\tau_k\right]} + \ds\frac
{\E_{\mb\pi^k}\left[\tau_k\langle\mb\ups_{\tau_k},c\rangle|\,G^k_{\tau_k}>\gamma(\eps)\right]}
{\E_{\mb\pi^k}\left[\tau_k\right]}
\P_{\mb\pi^k}\left(G^k_{\tau_k}>\gamma(\eps)\right)\\[15pt]
\le \gamma(\eps)+ g_{\max}\P_{\mb\pi^k}\left(G
^k_{\tau_k}>\gamma(\eps)\right) \,, \ea
$$
where $g_{\max}:=\max\left\{g(s,u)|\,s\in\mc S,u\in\mc U\right\}$.
From (\ref{PGtauk->0}) we get
$$
\limsup\limits_{k\in\N} \ds
\frac1{\E_{\mb\pi^k}\left[\tau_k\right]}
{\E_{\mb\pi^k}\left[\summ_{t=1}^{\tau_k}g(S_t,U_t)\right]} \le
\gamma(\eps)+g_{\max}\limsup\limits_{k\in\N}
\P_{\mb\pi^k}\left(G_{\tau_k}> \gamma(\eps)\right)=
\gamma(\eps)\,.
$$
Therefore (\ref{lemmac*claim}) follows from the arbitrariness of
$\eps>0$, and the fact that, as a consequence of Lemma
\ref{lemmasemicontinuity}, we have
$$\lim_{\eps\ra0^+}\gamma(\eps)=G\,.$$\qed
\subsection{Proof of Part 1 of Theorem \ref{maintheo}}
\label{sect4.4} We are now ready to step back to the problem of
upperbounding the error exponent of variable-length block-coding
schemes over FSMCs. We want to combine the result in Theorem
\ref{invtheo} with that in Theorem \ref{lemmac*} in order to
finally prove Part 1 of Theorem \ref{maintheo}.

Let $(\Phi^k,T_k,\Psi^k)$ be a sequence of variable-length block
coding schemes satisfying (\ref{constr}). Our goal is to prove
that \be \limsup_{k\in\N}\frac{-\log
p_e\left(\Phi^k,T_k,\Psi^k\right)}{\E_{\Phi^k}[T_k]}\le
D\left(1-\frac RC\right)\,. \label{goal}\ee A first simple
conclusion that can be drawn from Theorem \ref{invtheo}, using the
crude bounds $$c(\mb\Upsilon_{\Phi^k\!,t},S_t)\le\log|\mc
X|\,,\qquad d(S_t,\mb\Upsilon^i_{\Phi^k\!,t})\le d_{\max}\,,\
i=0,1\,,$$ is that \be \limsup\limits_{k\in\N} \ds\frac{-\log
p_e\left(\Phi^k,T_k,\Psi^k\right)}{\E_{\Phi^k}[T_k]}\le
\frac{D}{C}\log|\mc X|+ d_{\max}-R(1-\delta)
-\log\frac{\lambda\delta}4+\frac{D}{C}\ent(\delta)<+\infty\,.
\label{nomorethanexp}\ee Thus the error probability does not decay
to zero faster than exponentially with the expected transmission
time $\E_{\Phi^k}[T_k]$.

The core idea to prove (\ref{goal}) consists in introducing a real
sequence $(\delta_k)$ and showing that both
$$\tau_k:=\min\left\{T_k,\inf\left\{t\in\N\big|\,P_{MAP}^{\Phi^k}(t)\le\delta_k\right\}\right\}\,,$$
and $T_k-\tau_k$ diverge in the sense of satisfying
(\ref{probabilisticdivergence}).
The sequence $(\delta_k)$ needs to be carefully chosen: we want it
to be asymptotically vanishing in order to guarantee that $\tau_k$
diverges, but not too fast since otherwise $T_k-\tau_k$ would not
diverge. It turns out that one possible good choice is
$$\delta_k:=\frac{-1}{\log p_e\left(\Phi^k,T_k,\Psi^k\right)}\,.$$
It is immediate to verify that
$$\lim_{k\in\N}p_e\left(\Phi^k,T_k,\Psi^k\right)=0$$
implies \be \label{deltanreq}
\lim\limits_{k\in\N}\delta_k=0\,,\qquad
\lim\limits_{k\in\N}\frac{p_e\left(\Phi^k,T_k,\Psi^k\right)}{\delta_k}=0\,.
\ee
\begin{lemma}\label{lemmaP(T)}
In the previous setting, for every fixed $M$ in $\N$, we have
\be\lim\limits_{k\in\N}\P_{\Phi^k}\left(\tau_k\le
M\right)=0\,,\qquad\qquad
\lim_{k\in\N}\P_{\Phi^k}\left(T_k-\tau_k\le M\right)=0\,.
\label{Tn->inftyin Prob}\ee
\end{lemma}
\proof See Appendix \ref{AppendixB}.$\hfill\square$

Lemma \ref{lemmaP(T)} allows us to apply Theorem \ref{lemmac*}
first to the mutual information cost $c$ obtaining
$$
\limsup_{k\in\N}\frac{C_{\delta_k}(\Phi^k,T_k)}{\E_{\Phi^k}[\tau_k]} =
\limsup_{k\in\N}\frac{\E_{\Phi^k}\left[\summ_{t=1}^{\tau_k}c(S_t,\mb\Upsilon_{\Phi,t})\right]}
{\E_{\Phi^k}[\tau_k]}\le C
$$
and then to the information divergence cost $d$ obtaining
$$
\ds\limsup_{k\in\N}\frac{D_{\delta_k}(\Phi_k,T_k)}{\E_{\Phi^k}[T_k-\tau_k]}\le D\,.
$$
Therefore, by applying Theorem \ref{invtheo} we get
$$
\ba{rcl}\ds
D&\ge&\limsup\limits_{k\in\N}\ds\frac1{\E_{\Phi^k}[T_k]}
\left(\frac DC C_{\delta_k}(\Phi^k,T_k)
+D_{\delta_k}(\Phi^k,T_k)\right)\\[15pt]
&\ge& \limsup\limits_{k\in\N}\ds \frac{-\log
p_e(\Phi^k,T_k,\Psi^k)}{\E_{\Phi^k}[T_k]} +\frac DC\frac{\log|\mc
W_k|}{\E_{\Phi^k}[T_k]}(1-\alpha_k)+
\frac{\beta_k}{\E_{\Phi^k}[T_k]}
\\[15pt]
&=&\ds \frac DCR+ \limsup\limits_{k\in\N}\frac{-\log
p_e(\Phi^k,T_k,\Psi^k)}{\E_{\Phi^k}[T_k]} \ea$$ thus proving
(\ref{maintheo1}).

\section{An asymptotically optimal scheme}
\label{sect5}
\begin{figure}
\centering
\includegraphics[height=3.0cm,width=12cm]{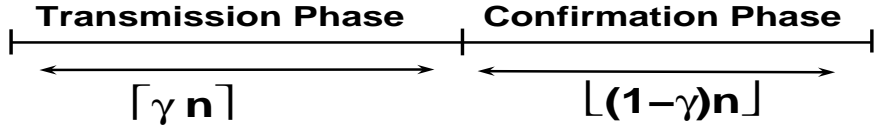}
\caption{One epoch in the generalized Yamamoto-Itoh scheme: a
total length $n$ is divided into two phases: a transmission one of
length $\hat n=\lceil\gamma n\rceil$ and a confirmation one of
length \mbox{$\hat n=\lfloor(1-\gamma) n\rfloor$.}}
\label{figureYamaItoh}
\end{figure}

In this section we propose and analyze a family of causal coding
schemes with feedback asymptotically achieving the Burnashev
exponent $E_B(R)$, thus proving Part 2 of Theorem \ref{maintheo}.

The scheme we propose can be viewed as a generalization of the one
introduced by Yamamoto and Itoh in \cite{YamamotoItoh} and
consists of a sequence of epochs.  Each epoch is made up of two
distinct fixed-length transmission phases, respectively named
communication and confirmation phase. In the communication phase
the message to be sent is encoded in a block code and transmitted
over the channel. At the end of this phase the decoder makes a
tentative decision about the message sent based on the observation
of the channel outputs and of the state sequence. As perfect
causal feedback is available at the encoder, the result of this
decision is known at the encoder. In the confirmation phase a
binary acknowledge message, confirming the decoder's estimation if
it is correct, or denying it when it is wrong, is sent by the
encoder through a fixed-length repetition code-function. The
decoder performs a binary hypothesis test in order to decide
whether a deny or a confirmation message has been sent. If an
acknowledge is detected the transmission halts, while if a deny is
detected the system restarts transmitting the same message with
the same protocol. Again because of perfect feedback availability
at the encoder, there are no synchronization problems.

More precisely we design our scheme as follows. Given a design
rate $R$ in $(0,C)$, let us fix an arbitrary $\gamma$ in $(\frac
RC,1)$. For every $n$ in $\N$, consider a message set $\mc W_n$ of
cardinality $|\mc W_n|=\exp(\lfloor n R\rfloor)$ and two
blocklengths $\hat n$ and $\tilde n$ respectively defined as $\hat
n=\lceil n\gamma\rceil$, $\tilde n:=n-\hat n$.
\medskip

\noindent{\bf Fixed-length block-coding for the transmission phase}\\
It is known from previous works (see \cite{TatikondaMitter} for
instance) that the capacity $C$ of the stationary Markov channel
we are considering is achievable by fixed-length block-coding
schemes. Thus, since the rate of the first transmission phase is
below capacity,
$$\hat R:=\lim_{n\in\N}\frac{\log|\mc W_n|}{\hat n}=\frac R{\gamma}<C\,,$$
there exists a family of causal encoders $(\hat\Phi^n)$
parametrized by an index $n$ in $\N$
$$\hat\Phi^n=\left(\mc W_n,(\hat\phi^n_t)\right)\,,\qquad
\hat\phi^n_t:\mc W_n\times\mc S^t\times\mc Y^{t-1}\ra\mc X\,,$$
and a corresponding family $(\hat\Psi^n)$ of decoders of fixed
length $\hat n$ (notice that $n$ is  the index while $\hat n$ is
the blocklength)
$$\hat\Psi^n:\mc S^{\hat n}\times\mc Y^{\hat n}\ra\mc W_n\,,$$
with error probability asymptotically vanishing in $n$. More
precisely, since the state space $\mc S$ is finite, the pair
$(\hat\Phi^n,\hat\Psi^n)$ can be designed in such a way that the
probability $\P_{\hat\Phi^n}(\hat\Psi^n\ne W|W=w,S_1=s)$ of error
conditioned on the transmission of any message $w$ in $\mc W_n$
and of an initial state $s$ approaches zero uniformly with respect
both to $w$ and $s$, i.e. \be p(n):=\max_{w\in\mc
W_n}\max_{s\in\mc S}\P_{\hat\Phi^n}\left(\hat\Psi^n\ne
W\big|\,W=w,S_1=s\right)
\stackrel{n\ra\infty}{\longrightarrow}0\,.\label{p(n)->0}\ee The
triple $(\hat\Phi^n,\hat n,\hat\Psi^n)$ will be used in the first
phase of each epoch of our iterative transmission scheme.
\hfill$\square$
\medskip

\noindent{\bf Binary hypothesis test for the confirmation phase}\\
For the second phase, instead, we consider a causal binary input
encoder $\tilde\Phi^n$ based on the optimal stationary policies in
the maximization problem (\ref{Dx0x1}). More specifically, we
define $\tilde\Phi^n$ by
$$\tilde\phi^n_t:\{0,1\}\times\mc S^t\ra\mc X\,,\qquad\qquad
\tilde\phi^n_t(m,\mb s)=f^*_m(s_t)\,,\qquad m=0,1\,,\qquad
t=1,\ldots,\tilde n\,,$$ where $f^*_0,f^*_1:\mc S\rightarrow\mc X$
are such that
$$D=\summ_{s\in\mc S}\mb\mu_{f_0}(s)D\left(P(\,\cdot\,,\,\cdot\,|s,f_0(s))||P(\,\cdot\,,\,\cdot\,|s,f_1(s))\right)$$

Suppose that an acknowledge message $m=0$ is sent. Then it is easy
to verify that the pair sequence $(S_{t+1},Y_t)_{t=1}^{\tilde n}$
forms a Markov chain over the space of the achievable channel
state output pairs \be \mc Z:=\big\{(s_+,y)\in\mc S\times\mc Y\st
\exists\,s\in\mc S,\exists\,x\in\mc X:\,P(s_+,y|\,s,x)>0\big\}
\label{Zdef}\ee with transition probability matrix
$$P_0=\Big(P_0(s_+,y|s,y_-):=P(s_+,y|s,f^*_0(s))\Big)\,.$$
Analogously, if a deny message $m=1$ has been sent, then
$(S_{t+1},Y_t)_{t=1}^{\tilde n}$ forms a Markov chain with
transition probability matrix
$$P_1=\Big(P_1(s_+,y|s,y_-):=P(s_+,y|s,f^*_1(s))\Big)\,.$$
It follows that a decoder for $\tilde\Phi^n$ is a binary
hypothesis test between two Markov chain hypothesis. Notice that
for both chains the transition probabilities $P_0(s_+,y|s,y_-)$
and $P_1(s_+,y|s,y_-)$ respectively do not depend on the second
component $y_{-}$ of the past state only, but only on its first
component $s$ as well as on the full future state $(s_+,y)$.

When the coefficient $D$ is finite, as a consequence of Assumption
\ref{mixingassumption} and (\ref{lambda>0}), we have that both the
stochastic matrices $P_0$ and $P_1$ are both irreducible over $\mc
Z$, with the invariant measure of $P_i$ given by
$$\tilde{\mb\mu}_i\in\mc P(\mc Z)\,,\qquad
\tilde{\mb\mu}_i(s_+,y):=\summ_{s\in\mc
S}\mb\mu_{f_i}(s)P(s_+,y|\,s,f_i(s))\,, \qquad i=0,1\,.
$$
Using binary hypothesis test results for irreducible Markov chains
(see \cite{Natarajan}and \cite[pagg.72-82]{DemboZeitouni}) it is
possible to show that a decoder
$$\tilde\Psi^n:\left(\mc S\times\mc Y\right)^{\tilde n-1}\ra\{0,1\}$$
can be chosen in such a way that, asymptotically in $n$, its type
1 error probability achieves the exponent (recall (\ref{Dx0x1}))
$$
\ba{rcl}\ds\summ_{z,z_+\in\mc
Z}\!\!\!\tilde{\mb\mu}_0(z)P_0(z_+|\,z)
\log\frac{\tilde{\mb\mu}_0(z)P_0(z_+|\,z)}{\tilde{\mb\mu}_0(z)P_1(z_+|\,z)}
\!\!\!\!&=&\!\!\!\!\!\! \ds\summ_{\substack{\\s,s_+\in\mc
S\\\vspace{4pt}y_-,y\in\mc Y}}
\!\!\!\tilde{\mb\mu}_0(s,y_-)P(s_+,y|\,s,f_0^*(s))
\log\frac{P(s_+,y|\,s,f_0^*(s))}{P(s_+,y|\,s,f_1^*(s))}
\\
&=&\!\!\!\ds\summ_{s\in\mc S}\mb\mu_0(s)
D\big(P(\,\cdot\,,\,\cdot\,|\,s,f_0^*(s))||\,P(\,\cdot\,,\,\cdot\,|\,s,f_1^*(s))\big)
\\
&=& D\ea$$ while its type one error probability is vanishing. More
specifically, since the state space is finite, we have that,
defining $p_i(n)$ as the maximum over all possible initial states
of the error probability of the pair $(\tilde\Phi^n,\tilde\Psi^n)$
conditioned on the transmission of a $'i'$ confirmation message,
i.e.
$$p_i(n):=\max_{s\in\mc S}\P_{\tilde{\Phi}}\left(\tilde\Psi^n
\left(\mb S_{2}^{\tilde n},\mb Y_1^{\tilde n-1}\right)\ne
i\big|\,W=i,\,S_1=s\right)\,, \qquad i=0,1\,,$$ we have
\be\lim_{n\in\N}p_0(n)=0\,,\qquad\lim_{n\in\N}\frac{-\log
p_1(n)}{\tilde n}=D\,. \label{p0(n)->0}\ee

When the coefficient $D$ is infinite, then the stochastic matrix
$P_0$ is irreducible over $\mc Z$, while there exists at least a
pair $z,z_+$ in $\mc Z$ such that $P_0(z_+|z_-)>0$ while
$P_1(z_+|z_-)=0$. It follows that a sequence of binary tests
$(\tilde\Psi^n)$, with $\tilde\Psi^n:\left(\mc S\times\mc
Y\right)^{\tilde n-1}\ra\{0,1\}$, can be designed such that \be
\lim_{n\in\N}p_0(n)=0\,,\qquad p_1(n)=0\,,\
n\in\N\,.\label{p1=0}\ee Such a family of tests is given for
instance by allowing $\Psi^n(\mb z)$ to equal 0 if and only if the
$(\tilde n-1)$-tuple $\mb z$ contains a symbol $z_-$ followed by a
$z_+$.

\hfill$\square$
\medskip

Once fixed $\hat\Phi^n$, $\hat\Psi^n$, $\tilde\Phi^n$ and
$\tilde\Psi^n$, the iterative protocol described above defines a
variable-length block-coding scheme $(\Phi^n,T_n,\Psi^n)$. As
mentioned above the scheme consists of a sequence of epochs, each
of fixed length $n$; in particular we have
$$T_n=n\tau_n\,,$$ where
$$\tau_n:=
\inf\left\{k\in\N:\tilde\Psi^n(\mb S_{(k-1)n+\hat n+1}^{kn},\mb
Y_{(k-1)n+\hat n+1}^{kn})=0\right\}\,,$$ is a positive integers
valued random variable describing the number of epochs occurred
until transmission halts.

The following result characterizes the asymptotic performances of
the family of schemes $(\Phi^n,T_n,\Psi^n)$.
\begin{proposition}For every design rate $R$ in $(0,C)$, and $\gamma$ in $(0,C)$,
we have \label{lemmaachievability} \be\lim_{n\in\N}\frac{\log|\mc
W_n|}{\E_{\Phi^n}[T_n]}=R\ee and
\begin{itemize}
\item if $D<+\infty$ \be\lim_{n\in\N}\frac{-\log
p_e(\Phi^n,T_n,\Psi^n)}{\E_{\Phi^n}[T_n]}=D(1-\gamma)\,,\label{achievgamma}\ee
\item if $D=+\infty$ \be p_e(\Phi^n,T_n,\Psi_n)=0\,,\qquad
n\in\N\,. \label{pe=0}\ee
\end{itemize}
\end{proposition}
\proof We introduce the following notation. First, for every
$k\in\N$:
\begin{itemize}
    \item $\hat e_k:=\{\hat\Psi(\mb S_{(k-1)n+1}^{(k-1)\tilde n},\mb Y_{(k-1)n+1}^{(k-1)\tilde n})\ne W\}$
          is the error event of the first transmission phase of the $k$-th epoch;
    \item $\tilde e_k:=\{\tilde\Psi(\mb S_{(k-1)n+\hat n+2}^{kn},\mb Y_{(k-1)n+\hat n+1}^{kn-1})\ne
           \1_{e_k}\}$
          is the error event of the second transmission phase of the $k$-th epoch;
\end{itemize}
Clearly we have
$$
\P_{\Phi^n}(\hat e_k|\mc F_{(k-1)n})\le p(n)\,,\qquad
\P_{\Phi^n}(\tilde e_k|\mc F_{(k-1)n+\hat n})\le
p_{\1_{e_k}}(n)\,.
$$
The transmission halts the first time a confirmation is detected
at the end of the second phase, i.e. the first time either a
correct transmission in the first phase is followed by a
successful transmission of an acknowledge message in the second
phase, or an uncorrect transmission in the first phase is followed
by a misreceived transmission of a deny message in the second
phase. It follows that we can rewrite $\tau_n$ as
$$
\tau_n=\inf\left\{k\in\N\st \left(e_k\cap\tilde e_k\right)\cup
\left(\left(e_k\right)^c\cap\left(\tilde
e_k\right)^c\right)\right\}\,.
$$
We claim that \be\P_{\Phi^n}(\tau_n\ge
k)\le\left(p(n)+p_0(n)\right)^{k-1}\,.\label{p(tau>=k)}\ee Indeed
(\ref{p(tau>=k)}) can be shown by induction. It is clearly true
for $k=1$. Suppose it is true for some $k$ in $\N$; then
$$\ba{rcl}
\P_{\Phi^n}(\tau_n\ge k+1)&=&\P_{\Phi^n}(\tau_n\ge k+1|\tau_n\ge k)\P_{\Phi^n}(\tau_n\ge k)\\[10pt]
&=& \big(\P_{\Phi^n}(e_{k+1})\left(1-\P_{\Phi^n}(\tilde
e_{k+1}|e_k)\right)+
\left(1-\P_{\Phi^n}(e_{k+1})\right)\P_{\Phi^n}(\tilde
e_{k+1}|\,(e_k)^c)\big)
\P_{\Phi^n}(\tau_n\ge k)\\[10pt]
&\le&
\left(p(n)+p_0(n)\right)\P_{\Phi^n}(\tau_n\ge k)\\[10pt]
&\le& \left(p(n)+p_0(n)\right)^{k}\,. \ea$$ Thus $\tau_n$ is
stochastically dominated by the sum of a constant $1$ plus a r.v.
with geometric distribution of parameter $p(n)+p_0(n)$. It follows
that its expected value can be bounded as follows
$$1\le\E_{\Phi^n}\left[\tau_n\right]=
\summ_{t\ge1}\P_{\Phi^n}\left(\tau_n\ge t\right)\le
\summ_{t\ge1}\left(p(n)+p_0(n)\right)^{t-1}\le
\frac1{1-\left(p(n)+p_0(n)\right)}\,.$$ Hence, from
(\ref{p(n)->0}) and (\ref{p0(n)->0}) we have
\be\lim_{n\in\N}\E_{\Phi^n}[\tau_n]=1\,.\label{E[tau]bound}\ee
From (\ref{E[tau]bound}) it immediately follows that
$$\lim_{n\in\N}\frac{\log|\mc W_n|}{\E_{\Phi^n}[T_n]}=
\lim_{n\in\N}\frac{\log\left(\exp(\lceil
nR\rceil)\right)}{n\E_{\Phi^n}[\tau_n]}=R\,.
$$
Moreover, transmission ends with an error if and only if an error
happens in the first transmission phase followed by a type-1 error
in the second phase, so that, the error probability of the overall
scheme $(\Phi^n,T_n,\Psi^n)$ can be bounded as follows \be\ba{rcl}
p_e(\Phi^n,T_n,\Psi^n)&=&\P_{\Phi^n}\left(e_{\tau_n}\cap \tilde e_{\tau_n}\right)\\[10pt]
&=&
\summ_{t\ge1}\P_{\Phi^n}\left(e_t\cap \tilde e_t\cap\{\tau_n= t\}\right)\\[10pt]
&=&
\summ_{t\ge1}\P_{\Phi^n}\left(e_t\cap \tilde e_t\cap\{\tau_n\ge t\}\right)\\[10pt]
&=& \summ_{t\ge1}\P_{\Phi^n}\left(e_t\cap\tilde e_t\big|\,\{\tau_n\ge
t\}\right)
\P_{\Phi^n}(\tau_n\ge t)\\[10pt]
&\le&
p(n)p_1(n)\summ_{t\ge1}\P_{\Phi^n}(\tau_n\ge t)\\[10pt]
&\le& \ds\frac{p(n)p_1(n)}{1-p(n)p_0(n)}\,. \ea\label{pebound}\ee
When $D$ is infinite, (\ref{pebound}) directly implies (\ref{pe=0}).
When $D$ is finite from (\ref{p(n)->0}), (\ref{p0(n)->0}), (\ref{E[tau]bound}) and
(\ref{pebound}) it follows that
$$
\ba{rcl} \liminf\limits_{n\in\N} \ds\frac{-\log
p_e(\Phi^n,T_n,\Psi^n)}{\E_{\Phi^n}[T_n]} &=&\ds
\liminf\limits_{n\in\N}
\ds\frac{-\log p_e(\Phi^n,T_n,\Psi^n)}{n\E_{\Phi^n}[\tau_n]}\\[10pt]
&\ge&\ds \liminf\limits_{n\in\N}
\left(1-p(n)p_0(n)\right)\left(\frac{-\log p_1(n)}{n} +
\frac{1}{n}\log\left(1-p(n)p_0(n)\right)\right)
\\[10pt]
&=& D\left(1-\gamma\right)\,, \ea
$$
which proves (\ref{achievgamma}).
\qed

It is clear that (\ref{maintheo2}) follows from
(\ref{achievgamma}) and the arbitrariness of $\gamma$ in $(\frac
RC,1)$ , so that Part 2 of Theorem \ref{maintheo} is proved.

We end this section with the following observation. It follows
from (\ref{p(tau>=k)}) that the probability that the proposed
transmission scheme halts after more than one epoch is bounded by
$p(n)+p_0(n)$, a term which is vanishing asymptotically with $n$.
Then, even if the transmission time is variable, it is constant
with high probability. As also observed in \cite{GallagerNakiboglu} in the memoryless
case, this is a desirable property from a practical point of view.

\section{An example} \label{sectexample}

\begin{figure}
\centering
\includegraphics[height=2.5cm,width=10cm]{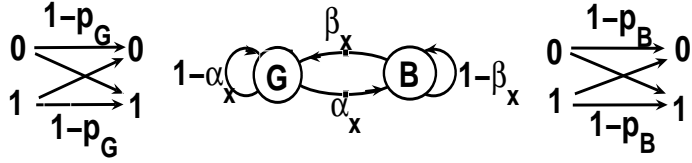}
\caption{A simple FSMC with binary state space $\mc S=\{G,B\}$ and
binary input/output space $\mc X=\mc Y=\{0,1\}$: notice that the
state transition probabilities are allowed to depend on the
current input (ISI)} \label{figureexample}
\end{figure}

We consider a FSMC as in Fig.\ref{figureexample}, with state space
$\mc S=\{G,B\}$, input and output spaces
 $\mc X=\mc Y=\{0,1\}$ and stochastic kernel given by:
     $$P(s_+,y|s,x)=P_S(s_+|x,s)P_Y(y|x,s)\,,\qquad s,s_+\in\mc S\,,\ x,y\in\mc\{0,1\}\,,$$
         $$P_S(B|G,0)=\alpha_0\,\qquad P_S(B|G,1)=\alpha_1\,\qquad P_S(G|B,0)=\beta_0\,\qquad P_S(G|B,1)=\beta_1\,, $$
         $$P_Y(1|G,0)=P_Y(0|G,1)=p_G\,,\qquad P_Y(1|B,0)=P_Y(0|B,1)=p_B\,,$$
where $0<p_G<p_B<\frac12$, and
$\alpha_0,\alpha_1,\beta_0,\beta_1\in(0,1)$.
For any stationary policy $\pi:\mc S\ra\mc P(\{0,1\})$, the state
invariant measure associated to $\pi$ can be made explicit:
$$\mb\mu_{\pi}(B)=\ds\frac{\alpha_0[\pi(G)](0)+\alpha_1[\pi(G)](1)}
{\alpha_0[\pi(G)](0)+\alpha_1[\pi(G)](1)+\beta_0[\pi(B)](0)+\beta_1[\pi(B)](1)}\,,
\qquad\mb\mu_{\pi}(G)=1-\mb\mu_{\pi}(B)\,.$$ The mutual
information costs are given by
$$c(G,u)=
\ent\left(u(1)\alpha_1+u(0)\alpha_0\right)+
\ent\left(u(1)p_G+u(0)(1-p_G)\right)-\ent(p_G)
-\left(u_G\ent(\alpha_1)+u(0)\ent(\alpha_0)\right)\,,$$
$$c(B,u)=
\ent\left(u(1)\beta_1+u(0)\beta_0\right)+
\ent\left(u(1)p_B+u(0)(1-p_B)\right)-\ent(p_B)
-\left(u(1)\ent(\beta_1)+u(0)\ent(\beta_0)\right)\,,$$ $\ent$
denoting the binary entropy function.
\begin{figure}
\centering
\includegraphics[height=7.0cm,width=7.0cm]{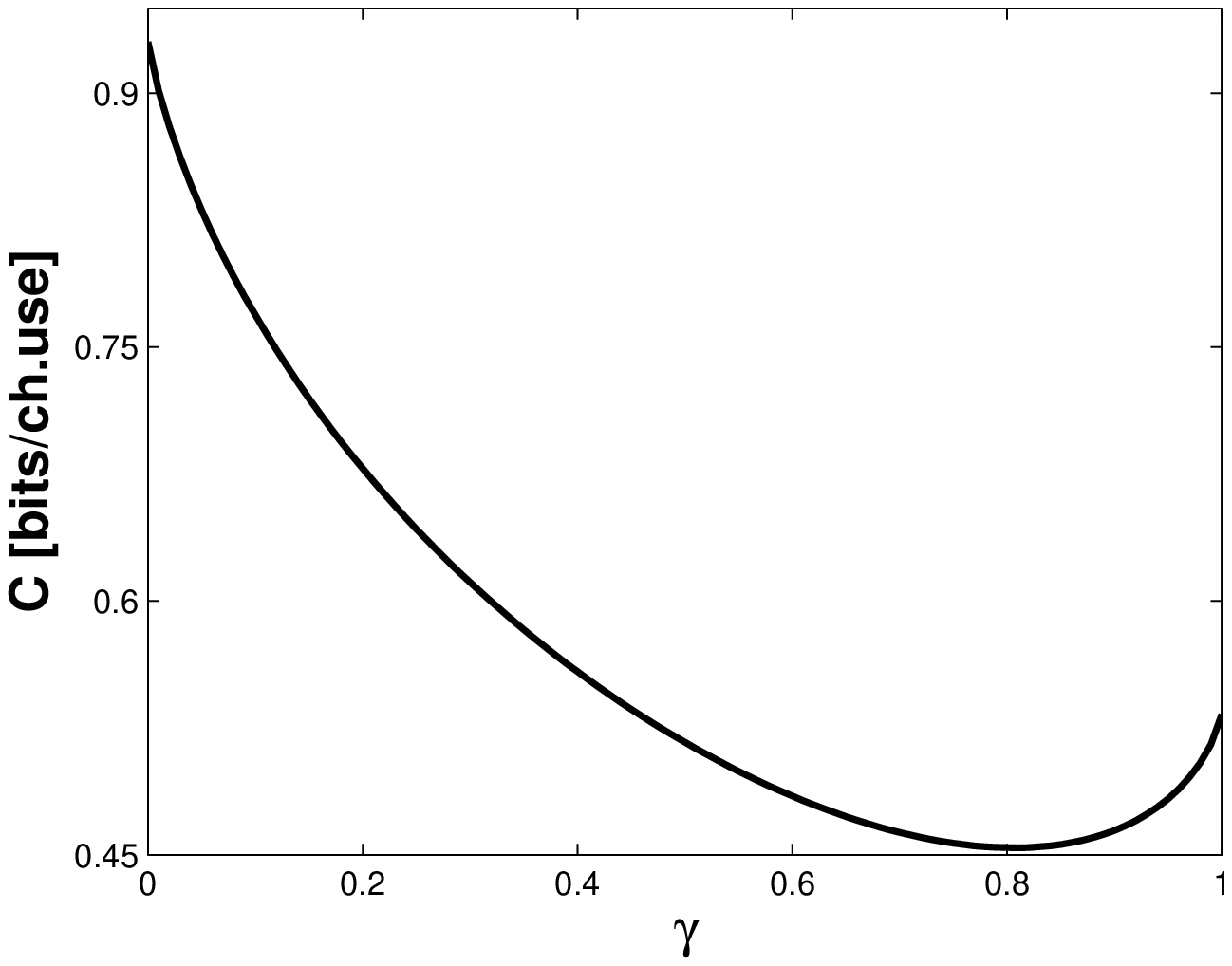}
\hspace{1cm}
\includegraphics[height=7.0cm,width=7.0cm]{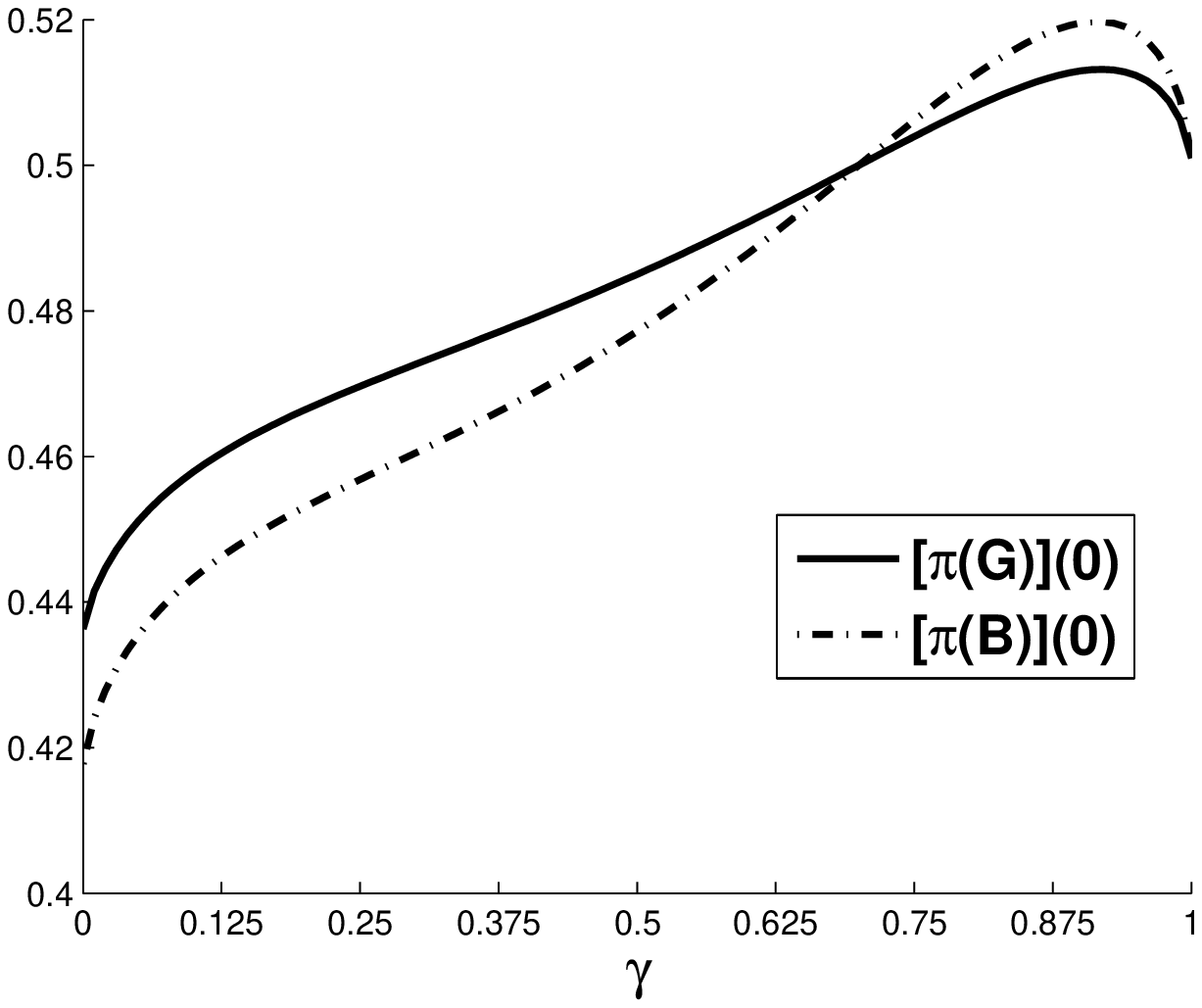}
\caption{In the righthand picture, the capacity of the FSMC in
Fig. \ref{figureexample} for values of the parameters $p_G=0.001$,
$p_B=0.1$, $\alpha_0=1-\beta_0=0.7$, $\alpha_1=1-\beta_1=\gamma$,
is plotted as a function of $\gamma$ in $(0,1)$. In the righthand
picture, for the same values of the parameters, the optimal policy
$\mb\pi^*:\{G,B\}\rightarrow\mc P(\{0,1\})$ is plotted as a
function of $\gamma$ in $(0,1)$.} \label{figurecapacity}
\end{figure}
The information divergence costs instead are given by
$$d\left(G,\delta_{f_0(G)}\right)=D\left(p_G||1-p_G)+D(\alpha_{f_0(G)}||\,\alpha_{f_1(G)}\right)
\,,$$$$
d(B,\delta_{x_B^0})=D\left(p_B||1-p_B)+D(\alpha_{f_0(G)}||\,\alpha_{f_1(G)}\right)\,,$$
where, for $x,y$ in $[0,1]$, $D(x||y):=x\log\frac
xy+(1-x)\log\frac{1-x}{1-y}$.
\begin{figure}
\centering
\includegraphics[height=7.0cm,width=7.0cm]{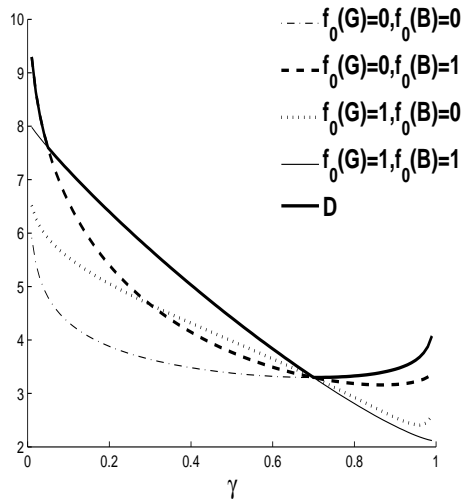}
\caption{The thick solid line is a plot of the Burnashev
coefficient $D$ (evaluated with natural log base) of the FSMC of
Fig.\ref{figureexample} for the same values of the parameters as
in Fig.\ref{figurecapacity}. } \label{figureDgamma}
\end{figure}

In Fig. \ref{figurecapacity} and Fig.\ref{figureDgamma} the
special case when $p_G=0.001$, $p_B=0.1$, $\alpha_0=1-\beta_0=0.7$
and $\alpha_1=1-\beta_1=\gamma$ is studied as a function of the
parameter $\gamma$ in $(0,1)$. In particular in
Fig.\ref{figurecapacity} the capacity and the optimal policy
$\pi:\mc S\ra\mc X$ are plotted as a function of $\gamma$. Notice
that for $\gamma=0.7$ the channel has no ISI and actually
coincides with a memoryless Gilbert-Elliot channel: for that value
the optimal policy chooses the uniform distribution both in the
good state $G$ as well as in the bad state $B$.
For values of $\gamma$ below $0.7$ (resp. beyond $0.7$), instead,
the optimal policy puts more mass on the input symbol 1 (resp. the symbol 0)
both in state $G$ and state $B$, and it is more unbalanced in state $B$.
In Fig.\ref{figureDgamma} the Burnashev coefficient of the channel is
plotted as a function of the parameter $\gamma$, as well as the
the values of the ergodic Kullback-Leibler cost corresponding to
the four possible policies $f_0:\{G,B\}\rightarrow\{0,1\}$.
Observe as the minimum value of $D$ is achieved for $\gamma=0.7$;
in that case all the four non trivial policies $f_0,f_1$ give the same
value of the Kullbak-Leibler cost.

Finally it is worth to consider the simple non-ISI case when
$\alpha_0=\alpha_1=\beta_0=\beta_1$. In this case the state ergodic measure
is the uniform one on $\{G,B\}$.
Notice by a basic convexity argument we get that its capacity $C$
and Burnashev coefficient $D$ satisfy
\be C=1-\frac12\ent(p_G)-\frac12\ent(p_B)>1-\ent(\frac12p_G+\frac12p_B)=:\tilde C\,,
\label{C>tilde C}\ee
\be D=\frac12D(p_G||1-p_G)+\frac12D(p_B||1-p_B)\ge
D(\frac12p_G+\frac12p_B||1-\frac12p_B-\frac12p_G)=:\tilde D\,.
\label{D>tildeD}\ee
In the (\ref{C>tilde C}) and (\ref{D>tildeD}) $\tilde C$ and $\tilde D$
correspond respectively to the capacity and the Burnashev coefficient of
memoryless binary symmetric channel with crossover probability equal
to the ergodic average of the crossover probabilities $p_B$ and $p_G$.
Such a channel is introduced in practice when channel interleavers
are used in order to apply to FSMCs coding techniques designed for DMCs.
While this approach reduces the decoding complexity, it is well known
that it reduces the achievable capacity (\ref{C>tilde C}) (see \cite{GoldsmithVaraya}).
Inequality (\ref{D>tildeD}) shows that this approach causes also
a loss in the Burnashev coefficient of the channel.

\section{Conclusions}
In this paper we studied the error exponent of FSMCs with
feedback. We have proved an exact single-letter characterization
of the reliability function for variable-length block-coding
schemes with perfect causal output feedback, generalizing the
result obtained by Burnashev \cite{Burnashev} for memoryless
channels. Our assumptions are that the channel state is causally
observable both at the encoder and the decoder and the stochatic
kernel describing the channel satisfies some mild ergodicity
properties.

As a first topic for future research, we would like to extend our
result to the case when the state is either observable
at the encoder only or it is not observable at neither side. We believe
that the techniques used in \cite{TatikondaMitter} in order to characterize
the capacity of FSMCs with state not observable may be adopted to
handle our problem as well. The main idea consists in studying a
partially observable Markov decision process and reduce it to a
fully observable one with a larger state space. However some
technical issues may appear since, in order to deal with average
cost problems, we used finiteness of the state space in our proofs
in Section \ref{sect3}.
Finally, it would be interesting to consider the problem of finding
universal schemes which do not require exact knowledge of the channel
statistics but use feedback in order to estimate them.

\appendix

\section{Proofs for Section 3}\label{AppendixA}
For the reader's convenience all statements are repeated before
their proof.
\setcounter{theorem}{1}
\begin{lemma}
Given any causal feedback encoder $\Phi$, we have, for every $t$
in $\N$,
$$\tilde P^{\Phi}_{MAP}(t)\ge \lambda\tilde P_{MAP}^{\Phi}(t-1)\qquad \P_{\Phi}-a.s.$$
\end{lemma}
\proof A first observation is that
$$
\P_{\Phi}\left(\cap_{x\in\mc
X}\big\{P(S_{t+1},Y_t|\,S_t,x)=0\big\}\right) =0\,,\qquad
\forall\, t\,\in\N\,.
$$
It follows that, $\P_{\Phi}$-almost surely, for every $t$ in $\N$
$$P\left(S_{t+1},Y_t|\,S_t,X_t\right)\ge\lambda_{S_t}\ge\lambda\,.$$

Let us fix an arbitrary message $w$ in $\mc W$. We have
$$
\ba{rcl} \P_{\Phi}\left(W=w|\,\mc G_{t}\right) &\ge&
\P_{\Phi}\left(W=w|\,\mc G_{t}\right)
\P_{\Phi}\left(S_{t+1},Y_t|\,\mc G_{t}\right)\\
[10pt] &=& \P_{\Phi}\left(W=w|\,\mc G_{t-1}\right)
\P_{\Phi}\left(S_{t+1},Y_t|\,W=w,\,\mc G_{t-1}\right)\\
[10pt] &=& \P_{\Phi}\left(W=w|\,\mc G_{t-1}\right)
P\left(S_{t+1},Y_t|S_t,X_t\right)\\
[10pt] &\ge& \lambda\,\P_{\Phi}\left(W=w|\,\mc G_{t-1}\right)\,.
\ea$$ It follows that
$$
\ba{rcl} \tilde P^{\Phi}_{MAP}(t)&=&
\P_{\Phi}\left(\tilde\Psi_{MAP}^t\ne W\big|\,\mc G_{t}\right)
\\[10pt]
&=& \summ_{\substack{w\in\mc W\\ w\ne\tilde\Psi^{t}_{MAP}}}
\P_{\Phi}(W=w|\mc G_{t})\\[25pt]
&\ge& \summ_{\substack{w\in\mc W\\ w\ne\tilde\Psi^{t}_{MAP}}}
\lambda\,\P_{\Phi}(W=w|\mc G_{t-1})
\\&\ge&
\lambda\, \tilde P^{\Phi}_{MAP}(t-1)\,. \ea
$$
\qed

\begin{lemma}
For any variable-length block-coding scheme
$\left(\Phi,T,\Psi\right)$ and any $0<\delta<\frac12$, we have $$
C_{\delta}(\Phi,T)\ge
\left(1-\delta-\frac{p_e\left(\Phi,T,\Psi\right)}{\delta}\right)
\log|\mc W|-\ent(\delta)\,. $$
\end{lemma}
\proof For every $n$ we introduce a random variable $\Gamma_n$
describing the conditional message entropy given the information
$\mc G_n$ available at the encoder at time $n+1$. Consider the
real valued random variable $V_n$ defined by
$$V_n:=\Gamma_{n}+\summ_{t=1}^nc\left(S_t,\mb\Upsilon_{\Phi,t}\right)
\,,\qquad n\in\Z^+\,,$$ We claim that $\left(V_n,\mc
G_{n}\right)_{n\in\Z^+}$ is a submartingale. Indeed, for every $n$
in $\Z^+$, $V_n$ is $\mc G_{n}$-measurable, since $\Gamma_{n}$ is,
and so do both $S_t$ and $\mb\Upsilon_{\Phi,t}$ for every $1\le
t\le n$. Moreover we have
$$
\ba{rcl} \E_{\Phi}\left[ \Gamma_{n-1}-\Gamma_{n}\big|\,\mc
G_{n-1}\right] &=&\ds \E_{\Phi}\left[\log\frac
{\P_{\Phi}\left(W|\,\mc G_{n}\right)} {\P_{\Phi}\left(W|\,\mc
G_{n-1}\right)}\big|\,\mc G_{n-1}\right]
\\[8pt]
&=&\ds \E_{\Phi}\left[\log\frac
{\P_{\Phi}\left(S_{n+1},Y_n|\,W,\mc G_{n-1}\right)}
{\P_{\Phi}\left(S_{n+1},Y_n|\,\mc G_{n-1}\right)} \big|\,\mc
G_{n-1}\right]
\\[8pt]
&\le&\ds \E_{\Phi}\left[\log\frac
{\P_{\Phi}\left(S_{n+1},Y_n|\,X_n,\mc G_{n-1}\right)}
{\P_{\Phi}\left(S_{n+1},Y_n|\,\mc G_{n-1}\right)} \big|\,\mc
G_{n-1}\right]
\\[8pt]
&=&\ds \summ_{x\in\mc X}\summ_{y\in\mc Y}\summ_{s_+\in\mc S}
\mb\Upsilon_{\Phi,n}(x)P(s_+,y|s,x)\log\frac{P(s_+,y|s,x)}
{\summ_{z\in\mc X}\mb\Upsilon_{\Phi,n}(z)P(s_+,y|s,z)}
\\[8pt]
&=& c\left(\mb\Upsilon_{\Phi,n},S_n\right) \,, \ea
$$
the inequality in the formula above following from the data
processing inequality once noted that, because of the causality of
the encoder and the Markovian structure of the channel,
$$\left(W,\mb S_1^n,\mb Y_1^{n-1}\right)\ - \
(X_{n},S_n)\ -\ (Y_n,S_{n+1})$$ forms a Markov chain. It follows
that
$$\E_{\Phi}\left[V_{n}-V_{n-1}\big|\,\mc G_{n-1}\right]
= \E_{\Phi}\left[\Gamma_{n}-\Gamma_{n-1}+
c\left(\mb\Upsilon_{\Phi,t},S_t\right)\big|\,\mc G_{n-1}\right]
\ge0\,.$$ Moreover, $(V_n)$ has uniformly bounded increments since
$$|V_{n}-V_{n-1}|\le |c\left(\mb\Upsilon_{\Phi,t},S_t\right)|+|\Gamma_{n}-\Gamma_{n-1}|\le
\log|\mc X|+2\log|\mc W|<+\infty\,.$$

Doob's optional sampling theorem can thus be applied to the
submartingale $\left(V_n,\mc G_{n}\right)_{n\in\Z^+}$ and the
stopping time $\tau_{\delta}$, concluding that \be \log|\mc
W|=\E_{\Phi}[\Gamma_0|\,\mc G_0]= \E_{\Phi}\left[V_0|\,\mc
G_0\right] \le \E_{\Phi}\left[V_{\tau_{\delta}}\right] =
\E_{\Phi}\left[\Gamma_{\tau_{\delta}}\right]+
\E_{\Phi}\left[\summ_{t=1}^{\tau_{\delta}}c\left(S_t,\mb\Upsilon_{\Phi,t}\right)\right]\,.
\label{lemmaI}\ee Finally, combining (\ref{lemmaI}) with
(\ref{E[h]<=}), we obtain
$$
C_{\delta}(\Phi,T)=
\E_{\Phi}\left[\summ_{t=1}^{\tau_{\delta}}
c\left(S_t,\mb\Upsilon_{\Phi,t}\right)\right] \ge
\left(1-\delta-\frac{p_e\left(\Phi,T,\Psi\right)}{\delta}\right)
\log|\mc W|-\ent(\delta)\,.$$\qed

\begin{lemma}
Let $\tau$ and $T$ be stopping times for the filtration $\mc G$
such that $\tau\le T$, and consider a partition of the message set
as in (\ref{partition}). Then $$
L_i\le\E_{\Phi}\left[\summ_{t=\tau+1}^{T}
d\left(\mb\Upsilon_{\Phi,t}^i,S_t\right) \Big|\,W\in\mc W_i,\mc
G_{\tau}\right]\,,\qquad\P_{\Phi}-a.s.\,,\ i=0,1\,.$$
\end{lemma}
\proof We will prove the claim for $i=0$. Define for $t\ge0$
$$Z_t:=
\log\frac {\summ_{x\in\mc X}P(S_{t+1},Y_t|\,S_t,x)
\mb\Upsilon_{\Phi,t}^0(x)} {\summ_{x\in\mc
X}P(S_{t+1},Y_{t}|\,S_{t},x) \mb\Upsilon_{\Phi,t}^1(x)} \,,$$ with
the agreement $\log\frac00=0$. We have that \be
|Z_t|\le2\log\frac1{\lambda}\,. \label{|Zt|<=} \ee Indeed if
$P(S_{t+1},Y_t|\,S_t,x)=0$ for every $x$ in $\mc X$, then $Z_t=0$
by definition. If instead there exists $x$ in $\mc X$ such that
$P(S_{t+1},Y_t|\,S_t,x)>0$, then
$$\ba{rcl}|Z_t|&=&\ds\left|\log\frac
{\summ_{x\in\mc X}P(S_{t+1},Y_t|S_t,x) \mb\Upsilon_{\Phi,t}^0(x)}
{\summ_{x\in\mc X}P(S_{t+1},Y_{t}|S_{t},x)
\mb\Upsilon_{\Phi,t}^1(x)}\right|\\[15pt]
&\le& \ds2\log\left(\inf_{x\in\mc
X}\{P(S_{t+1},Y_t|S_t,x)\}\right)^{-1} =2\log\frac1{\lambda_{S_t}}
\le2\log\frac1{\lambda}\ea
$$

It is easy to check by induction that for every $n\ge0$ \be
\ds\log\frac{\P_{\Phi}\left(\mb S_{1}^{n+1},\mb
Y_{1}^{n}|\,W\in\mc W_0\right)} {\P_{\Phi}\left(\mb
S_{1}^{n+1},\mb Y_{1}^{n}|\,W\in\mc W_1\right)}
=\summ_{t=1}^nZ_t\,. \label{VnZt}\ee Indeed (\ref{VnZt}) holds
true for $n=0$, since $S_1$ is independent from $W$ (with the
agreement for an empty summation to equal zero). Moreover, suppose
(\ref{VnZt}) holds true for some $n$. Then
$$
\ba{rcl} \ds\log\frac {\P_{\Phi}\left(\mb S_1^{n+2},\mb
Y_1^{n+1}\big|\,W\in\mc W_0\right)} {\P_{\Phi}\left(\mb
S_1^{n+2},\mb Y_1^{n+1}\big|\,W\in\mc W_1\right)} &=&\ds \log\frac
{\P_{\Phi}\left(\mb S_1^{n+1},\mb Y_1^n\big|\,W\in\mc W_0\right)
\P_{\Phi}\left(S_{n+2},Y_{n+1}\big|\,W\in\mc W_0,\mc
E_{n+1}\right)} {\P_{\Phi}\left(\mb S_1^{n+1},\mb
Y_1^n\big|\,W\in\mc W_1\right)
\P_{\Phi}\left(S_{n+2},Y_{n+1}\big|\,W\in\mc W_1,\mc E_{n+1}\right)}\\
&=&\ds \log\frac {\P_{\Phi}\left(\mb S_1^{n+1},\mb
Y_1^n\big|\,W\in\mc W_0\right) \summ_{x\in\mc
X}P(S_{n+2},Y_{n+1}|S_{n+1},x) \mb\Upsilon_{\Phi,n+1}^0(x)}
{\P_{\Phi}\left(\mb S_1^{n+1},\mb Y_1^n\big|\,W\in\mc W_1\right)
\summ_{x\in\mc X}P(S_{n+2},Y_{n+1}|S_{n+1},x)
\mb\Upsilon_{\Phi,n+1}^1(x)}\\
&=&\ds \log\frac {\P_{\Phi}\left(\mb S_1^{n+1},\mb
Y_1^n\big|\,W\in\mc W_0\right)} {\P_{\Phi}\left(\mb S_1^{n+1},\mb
Y_1^n\big|\,W\in\mc W_1\right)} +Z_{n+1}=\summ_{t=1}^{n+1}Z_{t}\,.
\ea
$$

Now, by applying the log-sum inequality and recalling the
definition (\ref{ddef}) of the cost $d$, we have, for ever
$t\ge1$, \be \ba{rcl} \E_{\Phi}[Z_t|\,W\in\mc W_0,\mc E_{t}]&=&
\ds\E_{\Phi}\left[\log\frac{\summ_{x\in\mc X}P(S_{t+1},Y_t|S_t,x)
\mb\Upsilon_{\Phi,t}^0(x)} {\summ_{x\in\mc
X}P(S_{t+1},Y_{t}|S_{t},x)
\mb\Upsilon_{\Phi,t}^1(x)}\Bigg|\,W\in\mc W_0,\mc E_{t}\right]\\
&=& \ds\summ_{y\in\mc Y}\summ_{s\in\mc S}\left(\summ_{x\in\mc X}
\mb\Upsilon_{\Phi,t}^0(x)P(s,y|\,S_t,x)\right)
\log\frac{\summ_{x\in\mc
X}P(s,y|\,S_t,x)\mb\Upsilon_{\Phi,t}^0(x)}
{\summ_{x\in\mc X}P(s,y|\,S_t,x)\mb\Upsilon_{\Phi,t}^1(x)}\\
&\le& \ds\summ_{y\in\mc Y}\summ_{s\in\mc S}\left(\summ_{x\in\mc X}
\mb\Upsilon_{\Phi,t}^0(x)P(s,y|\,S_t,x)\right)
\log\frac{P(s,y|\,S_t,x)\mb\Upsilon_{\Phi,t}^0(x)}
{P(s,y|S_t,x)\mb\Upsilon_{\Phi,t}^1(x)}
\\[20pt]
&=&d(\mb\Upsilon_{\Phi,t}^0,S_t)\,. \ea \label{ZtK}\ee From
(\ref{VnZt}) and (\ref{ZtK}) it follows that, if we define
$$
V_n:= \log\frac {\P_{\Phi}\left(\mb S_1^{n+1},\mb
Y_1^n\big|\,W\in\mc W_0\right)} {\P_{\Phi}\left(\mb S_1^{n+1},\mb
Y_1^n\big|\,W\in\mc W_1\right)}
-\summ_{t=1}^nd(S_t,\mb\Upsilon_{\Phi,t}^0)\,,\qquad n\ge0\,,
$$
then $\left(V_n,\mc G_{n}\right)_{n\ge0}$ is a submartingale with
respect to the conditioned probability measure
$\P_{\Phi}(\,\cdot\,|\,W\in\mc W_0)$. Moreover it follows from
(\ref{|Zt|<=}) (recall that we are assuming $\lambda>0$ and that
this is equivalent to the boundedness of $K$) that $(V_n)$ has
uniformly bounded increments:
$$|V_{n+1}-V_n|\le|Z_{n+1}|+|d(S_t,\mb\Upsilon_{\Phi,n+1}^0)|
\le\log\frac1{\lambda}+d_{\max}<+\infty\,.$$ Thus, since $\tau\le
T$, Doob's optional stopping theorem can be applied yielding \be
\E_{\Phi}\left[V_{T}-V_{\tau}\big|\,W\in\mc W_0,\mc
G_{\tau}\right]\le0\,. \label{quasiclaim}\ee
Then the claim
follows from (\ref{quasiclaim}), after noticing that
$$V_{T}-V_{\tau}=
\log\frac {\P_{\Phi}\left(\mb S_{\tau+2}^{T+1},\mb
Y_{\tau+1}^{T}\big|\,W\in\mc W_0,\mc G_{\tau}\right)}
{\P_{\Phi}\left(\mb S_{\tau+2}^{T},\mb
Y_{\tau+1}^{T}\big|\,W\in\mc W_1,\mc G_{\tau}\right)}
-\summ_{t=\tau+1}^{T}d(S_t,\mb\Upsilon_{\Phi,t}^0)\,, \qquad
\P_{\Phi}-a.s.\,.$$ \qed

\begin{lemma}
Let $\Phi$ be any causal encoder, and $\tau$ and
$T$ be stopping times for the filtration $\mc G$ such that
$\tau\le T$. Then, for every $2^{\mc W}$-valued $\mc
G_{\tau}$-measurable r.v. $\mc W_1$, we have $\P_{\Phi}$-a.s.
$$
\E_{\Phi}\left[\summ_{t=\tau+1}^Td\left(S_t,\mb\Upsilon^{\1_{\{W\in\mc W_1\}}}_{\Phi,t}\right)
\big|\,\mc G_{\tau}\right] \ge
\log\frac{Z}{4}-\log\P\left(\tilde\Psi\ne\1_{\mc W_1}(W)
\big|\,\mc G_{\tau}\right)\,,$$ where
$$Z:=\min\Big\{
\P_{\Phi}\left(W\in\mc W_0|\,\mc G_{\tau}\right)\,,\
\P_{\Phi}\left(W\in\mc W_1|\,\mc G_{\tau}\right) \Big\}\,.$$
\end{lemma}
\proof First we will prove the statement when $\mc W_1$ is a
fixed, non-trivial subset of the message set $\mc W$. From the
log-sum inequality it follows that
$$
\ba{rcl} L_0&=& \ds\E_{\Phi}\left[\log\frac {\P_{\Phi}\left(\mb
S_{\tau+2}^{ T+1},\mb Y_{\tau+1}^{ T}\big|\,W\in\mc W_0,\mc
G_{\tau}\right)} {\P_{\Phi}\left(\mb S_{\tau+2}^{ T+1},\mb
Y_{\tau+1}^{ T}\big|\,W\in\mc W_1,\mc G_{\tau}\right)}\Bigg|\,
W\in\mc W_0,\mc G_{\tau}\right]
\\[7pt]&\ge&\ds
\summ_{i=0,1} \P_{\Phi}\left(\tilde\Psi=i\big|\,W\in\mc W_i,\mc
G_{\tau}\right) \log\frac
{\P_{\Phi}\left(\tilde\Psi=i\big|\,W\in\mc W_i,\mc
G_{\tau}\right)} {\P_{\Phi}\left(\tilde\Psi\ne i\big|\,W\notin\mc
W_i,\mc G_{\tau}\right)}
\\[7pt]\ds
&\ge& - \ent\left(\P_{\Phi}(\tilde\Psi=1\big|\,W\in\mc W_0,\mc
G_{\tau} )\right) -\P_{\Phi}\left(\tilde\Psi=0\big|\,W\in\mc
W_0,\mc G_{\tau}\right)
\log\P_{\Phi}\left(\tilde\Psi=1\big|\,W\in\mc W_0,\mc G_{\tau}\right)\\[7pt]
&\ge&\ds -\log 2 -\P_{\Phi}\left(\tilde\Psi=0\big|\,W\in\mc
W_0,\mc G_{\tau}\right)
\log\P_{\Phi}\left(\tilde\Psi=1\big|\,W\in\mc W_0,\mc
G_{\tau}\right)\,. \ea
$$
We now consider the error probability of $\tilde\Psi$ conditioned
on the sigma-field $\mc G_{\tau}$:
$$\ba{rcl}
\ds\P_{\Phi}\left(\tilde\Psi\ne\1_{\mc W_1}(W)\big|\,\mc
G_{\tau}\right) &=& \ds\P(W\in\mc W_0|\mc G_{\tau})
\P_{\Phi}(\tilde\Psi=1|\,W\in\mc W_0,\mc G_{\tau})\\
&&+\ds \P_{\Phi}(W\in\mc W_1|\mc G_{\tau})
\P_{\Phi}(\tilde\Psi=0|\,W\in\mc W_1,\mc G_{\tau})\\[5pt]
&\ge&\ds \min\limits_{i=0,1}\left\{\P_{\Phi}(W\in\mc W_i|\mc
G_{\tau})\right\} \P_{\Phi}(\tilde\Psi=1|\,W\in\mc W_0,\mc
G_{\tau})
\\[10pt]
&=&\ds Z\,\P_{\Phi}(\tilde\Psi=1|W\in\mc W_0,\mc G_{\tau})\,. \ea
$$
From Lemma \ref{lemmaD0} it follows that \be \ba{c}\ds
\ds \E_{\Phi}\left[\summ_{t=\tau+1}^{ T}
d\left(S_{t},\mb\Upsilon_{\Phi,t}^0\right)
\big|\,W\in\mc W_0,\mc G_{\tau}\right]\\
\ge\ds \E_{\Phi}\left[\log\frac {\P_{\Phi} \left(\mb
S_{\tau+1}^{ T},\mb Y_{\tau+1}^{ T}\big|\,W\in\mc W_0,\mc
G_{\tau}\right)} {\P_{\Phi}\left(\mb S_{\tau+1}^{ T},\mb
Y_{\tau+1}^{ T} \big|\,W\in\mc W_1,\mc G_{\tau}\right)}\big|\,
W\in\mc W_0,\mc G_{\tau}\right]
\\\ge\ds
-\log2-\P_{\Phi}\left(\tilde\Psi=0\big|\,W\in\mc W_0,\mc
G_{\tau}\right) \log\P_{\Phi}\left(\tilde\Psi=1\big|\,W\in\mc
W_0,\mc G_{\tau}\right)
\\\ge\ds
-\log2- \P_{\Phi} \left(\tilde\Psi=0\big|\,W\in\mc W_0,\mc
G_{\tau}\right) \log\left(\frac1Z\,
\P_{\Phi}\left(\tilde\Psi\ne\1_{\mc W_1}(W) \big|\,\mc
G_{\tau}\right)\right) \,.\ea \label{EK0}\ee An analogous
derivation leads to \be\label{EK1}
\ds \E_{\Phi}\left[\summ_{t=\tau+1}^{ T}
d\left(S_{t},\mb\Upsilon_{\Phi,t}^1\right)
\big|\,W\in\mc W_1,\mc G_{\tau}\right]\ge -\log2-
\P_{\Phi}\left(\tilde\Psi=1\big|\,W\in\mc W_1,\mc
G_{\tau}\right)\log \left(\frac1Z\, \P\left(\tilde\Psi\ne\1_{\mc
W_1}(W)\big|\,\mc G_{\tau}\right)\right) \,. \ee If we now average
(\ref{EK0}) and (\ref{EK1}) with respect to the posterior
distribution of $W$ given $\mc G_{\tau}$, we obtain (\ref{Wi<=K}).
Finally, since the claim holds true for every
choice of $\mc W_0$ in $2^{\mc W}\setminus\{\emptyset,\mc W\}$,
then it continues to hold true also when
$\mc W_0$ is a $2^{\mc W}\setminus\{\emptyset,\mc W\}$-valued $\mc
G_{\tau}$-measurable random variable. \qed

\begin{lemma}
Let $\Phi$ be a causal feedback encoder and $T$ a transmission time for $\Phi$.
Then, for every $0<\delta<1/2$ there exists a
$\mc G_{\tau_{\delta}}$-measurable random subset $\mc W_1$ of the
message set $\mc W$, whose a posteriori error probabilities
satisfy $$1-\lambda\delta\ge\P\left(W\in\mc W_i\big|\mc
G_{\tau_{\delta}}\right)\ge\lambda\delta\,,\qquad i=0,1\,.$$
\end{lemma}
\proof
Suppose first that
$\tilde P_{MAP}(\tau_{\delta})\le\delta$.
Then, since clearly $\tilde P_{MAP}(\tau_{\delta}-1)\ge\delta$, by
Lemma \ref{lemmaPMAP} we have
$$\tilde P_{MAP}(\tau_{\delta})\ge\lambda\,\tilde P_{MAP}(\tau_{\delta}-1)\ge\lambda\,\delta\,$$
It follows that if we define $\mc
W_1:=\{\Psi_{MAP}(\tau_{\delta})\}$, we have
$$
\P_{\Phi}(W\in\mc W_1|\,\mc G_{\tau_{\delta}})=
1-P_{MAP}(\tau_{\delta})\ge1-\delta\ge\lambda\,\delta\,, \qquad
\P_{\Phi}(W\notin\mc W_1|\,\mc G_{\tau_{\delta}})=
P_{MAP}(\tau_{\delta})\ge\lambda\,\delta\,.
$$
If instead $\tilde P_{MAP}(\tau_{\delta})>\delta$, the a
posteriori probability of any message $w$ in $\mc W$ at time
$\tau_{\delta}$ satisfies $\P_{\Phi}\left(W=w|\,\mc
G_{\tau_{\delta}}\right)\le1-\delta$. Then it is possible to
construct $\mc W_1$ in the following way. Introduce an arbitrary
labelling of $\mc W=\{w_1,w_2,\ldots,w_{|\mc W|}\}$. For any $1\le
i\le|\mc W|$, define $\mc W_{(i)}=\{w_1,\ldots,w_i\}$. Set
$k:=\inf\left\{1\le i\le|\mc W|\,:\, \P_{\Phi}\left(W\in\mc
W_{(i)}|\,\mc G_{t}\right)\ge\lambda\,\delta\right\}$, and define
$\mc W_1=\mc W_{(k)}$. Then clearly $\P_{\Phi}\left(W\in\mc
W_1|\,\mc G_t\right)\ge\lambda\,\delta$, while
$$\ba{rcl}\P_{\Phi}\left(W\notin\mc W_1|\,\mc G_{t}\right)&=&
1-\P_{\Phi}\left(W\in\mc W_{(k)}|\,\mc G_{t}\right)\\[8pt]
&=& 1-\P_{\Phi}\left(W\in\mc W_{(k-1)}|\,\mc G_{t}\right)
-\P_{\Phi}\left(W=w_k|\,\mc G_{t}\right)\\[8pt]
&\ge&1-\lambda\,\delta-(1-\delta)\ \ge\ \lambda\delta\,.\ea$$
\qed

\section{Proofs for Section 4}
\label{AppendixB}
\setcounter{theorem}{7}
\begin{lemma}
For every $\eps>0$, and for every feasible
policy $\mb\pi$ $$\P_{\mb\pi}\left(||\mb
F(\mb\ups_n)||\ge\eps+\frac1n\right)\le 2|\mc
S|\exp\left(-n\eps^2/2\right)\,. $$
\end{lemma}
\proof Let us fix an arbitrary admissible policy $\mb\pi$ in
$\Pi$. For every $s$ in $\mc S$ consider the following random
process:
$$Z_0^s:=0\,,\qquad Z_1^s:=0\,,$$
$$Z_n^s:=(n-1)F_s(\mb\ups_{n-1})+\1_{\{S_n=s\}}-\1_{\{S_1=s\}}\,,\qquad n\ge2\,.$$
We have
$$
\ba{rcl} Z_n^s&=&
(n-1)F_s(\mb\ups_{n-1})+\1_{\{S_n=s\}}-\1_{\{S_1=s\}}\\[10pt]
&=&(n-1)\mb\ups_{n-1}\left(\{s\},\mc U\right)
+\1_{\{S_n=s\}}-\1_{\{S_1=s\}} -(n-1)\int_{\mc S\times\mc
U}Q_S(s\,|\,j,u)\de\mb\ups_{n-1}(j,u)
\\[10pt]
&=&\summ_{t=2}^n\1_{\{S_t=s\}}-
\summ_{t=2}^nQ(s\,|\,S_{t-1},U_{t-1})\\[10pt]
&=&
\summ_{t=2}^n\left(\1_{\{S_t=s\}}-\E_{\pi}\left[\1_{\{S_t=s\}}|\mc
E_{t-1}\right]\right)\,. \ea
$$
It is immediate to check that $Z_n^s$ is $\mc E_n$-measurable.
Moreover
$$\E_{\mb\pi}[Z_{n+1}^s|\mc E_n]=Z_n^s\,,\forall\,n\ge0\,,$$
so that $\left(Z_n^s,\mc E_n,\P_{\mb\pi}\right)_{n\ge0}$ is a
martingale. Moreover, $(Z_n^s)$ has uniformly bounded increments
since $|Z_1^s-Z_0^s|=a_1:=0$, while
$$\left|Z_{n+1}^s-Z_n^s\right|=
\left|\1_{\{S_{n+1}=s\}}-\E_{\mb\mu}^{\pi}\left[\1_{\{S_{n+1}=s\}}|\mc
E_n\right]\right|\le a_{n+1}:=1\,, \qquad n\ge1\,.$$ It follows
that we can apply Hoeffding-Azuma inequality \cite{concentration},
obtaining
$$\P_{\mb\pi}\left(|Z_{n+1}^s|\ge\eps n\right)\le2\exp\left(-\frac{\eps^2n^2}{2\sum_{k=1}^{n+1}a_k}\right)
=2\exp\left(-\frac{\eps^2}2n\right)\,.$$ By simply applying a
union bound, we can conclude that
$$
\ba{rcl} \P_{\mb\pi}\left(||\mb
F(\mb\ups_n)||\ge\eps+\frac1n\right)&=&
\P_{\mb\pi}\left(\max_{s\in\mc S}\left|Z_{n+1}^s+\1_{\{S_1=s\}}-\1_{\{S_{n+1}=s\}}\right|\ge\eps n+1\right)\\[10pt]
&\le&
\P_{\mb\pi}\left(\bigcup_{s\in\mc S}\{\left|Z_{n+1}^s\right|\ge\eps n\}\right)\\[10pt]
&\le& \summ_{s\in\mc
S}\P_{\mb\pi}\left(\left|Z_{n+1}^s\right|\ge\eps n\right) \le
2|\mc S|\exp\left(-\frac{\eps^2}{2}n\right)\,. \ea$$ \qed

\begin{lemma}
The map $\gamma$ is upper semicontinuous. (i.e. $x_n\ra
x\Rightarrow\limsup_n\gamma(x_n)\le\gamma(x)$)
\end{lemma}
\proof
Possibly up to a subsequence,
with no loss of generality we can assume that
$$\gamma(x_n)\rightarrow\limsup_n\gamma(x_n)\,.$$
Since $\mc S\times\mc U$ is compact, the Prohorov space
$\mc P(\mc S\times\mc U)$ is compact as well \cite{Borkarbook}.
Thus, since the map $\mb\eta\mapsto||\mb F(\mb\eta)||$ is continuous,
the sublevel $\{||\mb F(\mb\eta)||\le x\}$ is compact.
It follows that for every $n$ there exists $\mb\eta_n$ in $\mc P(\mc S\times\mc U)$
such that
$$\gamma(x_n)=\sup\left\{\langle\mb\eta,g\rangle\big|\,\mb\eta\in\mc
P(\mc S\times\mc U):\, ||\mb F(\mb\eta)||\le x_n\right\}=\langle\mb\eta_n,g\rangle\,,\qquad
\mb ||F(\mb\eta_n)||\le x_n\,.
$$
Since $\mc P(\mc S\times\mc U)$ is compact we can extract a converging subsequence $(\eta_{n_k})$;
define $\ov{\mb\eta}:=\lim_k\mb\eta_{n_k}$.
Clearly
$$||\mb F(\ov{\mb\eta})||=\lim_k||\mb F(\mb\eta_{n_k})||\le x\,.$$
It follows that
$$\gamma(x)=
\sup\left\{\langle\mb\eta,g\rangle\big|\,\mb\eta\in\mc
P(\mc S\times\mc U):\, ||\mb F(\mb\eta)||\le x\right\}
\ge\langle\ov{\mb\eta},g\rangle=\lim_k\langle\mb\eta_{n_k},g\rangle=
\limsup_n\gamma(x_n)\,.
$$
\qed

\begin{lemma}
Let $(\tau_k)$ be a sequence of stopping times for the filtration
$\mc F$ and $(\mb\pi^k)$ be a sequence of feasible policies such
that $\E_{\mb\pi^k}[\tau_k]<\infty$ for every $k$ and
(\ref{probabilisticdivergence}) holds true. Then $$
\lim_{k\in\N}\P_{\mb\pi^k}\left(G^k_{\tau_k}>\gamma(\eps)\right)=0\,,\qquad
\forall\eps>0\,. $$
\end{lemma}
\proof For every $m$ in $\Z_+$ such that
$\P_{\mb\pi^k}\left(\tau_k\ge m\right)>0$ we have
$$
\ba{rcl}\P_{\mb\pi^k}\left(G^k_{\tau_k}>\gamma(\eps)|\,\tau_k\ge
m\right) &=& \summ_{t\ge m}
\ds\frac{\P_{\mb\pi^k}\left(\tau_k=t\right)}
{\P_{\mb\pi^k}\left(\tau_k\ge m\right)}
\P_{\mb\pi^k}\left(G^k_{\tau_k}>\gamma(\eps)|\,\tau_k=t\right)\\[10pt]
&\le& \summ_{t\ge m} \P_{\mb\pi^k}\left(\tau_k=t\right)
\P_{\mb\pi^k}\left(G^k_{\tau_k}>\gamma(\eps)|\,\tau_k=t\right)\\[10pt]
&\le& \summ_{t\ge0} \P_{\mb\pi^k}\left(\tau_k=t\right)
\P_{\mb\pi^k}\left(G^k_{\tau_k}>\gamma(\eps)|\,\tau_k=t\right)\\[10pt]
&=& \P_{\mb\pi^k}\left(G^k_{\tau_k}>\gamma(\eps)\right)\,. \ea
$$
An application of the Bayes rule thus gives us
$$
\P_{\mb\pi^k}\left(\tau_k\ge m\right)\ge
\P_{\mb\pi^k}\left(\tau_k\ge
m\big|\,G^k_{\tau_k}>\gamma(\eps)\right)\,,\qquad \forall\, k
\text{ s.t.
}\P_{\mb\pi^k}\left(G^k_{\tau_k}>\gamma(\eps)\right)>0\,,
$$
which in turns implies \be\ba{rcl}
\E_{\mb\pi^k}\left[\tau_k\right]&=&
\summ_{m\ge0}\P_{\mb\pi^k}\left(\tau_k\ge m\right)\\[10pt]
&\ge& \summ_{m\ge0}\P_{\mb\pi^k}\left(\tau_k\ge
m\big|\,G^k_{\tau_k}>\gamma(\eps)\right)=
\E_{\mb\pi^k}\left[\tau_k\big|\,G^k_{\tau_k}>\gamma(\eps)\right]\,.\ea
\ee On the other hand, for every $\eps>0$, using a union bound
estimation and (\ref{Azuma1}) we get, \be \ba{rcl}
\P_{\mb\pi^k}\left(G^k_n>\gamma\big(\eps+\frac1n\big)\right)&=&
\P_{\mb\pi^k}\left(\bigcup_{t\ge
n}\left\{\langle\mb\ups_t,c\rangle>
\gamma\big(\eps+\frac1n\big)\right\}\right)\\
&\le& \summ_{t\ge n}
\P_{\mb\pi^k}\left(\langle\mb\ups_t,c\rangle>\gamma\big(\eps+\frac1n\big)\right)\\
&\le&
2|\mc S|\summ_{t\ge n}\exp\left(-t\eps^2/2\right)\\
&=& 2|\mc S|\ds\frac{\exp\left(-n\eps^2/2\right)}
{1-\exp\left(-\eps^2/2\right)} \ea \label{Pk(G_n)<=exp(-n)} \ee It
follows that for every $M$ in $\N$ we have
$$
\ba{rcl}
\P_{\mb\pi^k}\left(G^k_{\tau_k}>\gamma\big(\eps+\frac1M\big)\right)&=&
\P_{\mb\pi^k}\left(\left\{G^k_{\tau_k}>
\gamma\big(\eps+\frac1M\big)\right\}\cap\{\tau_k\ge M\}\right) +
\P_{\mb\pi^k}\left(\left\{G^k_{\tau_k}>
\gamma\big(\eps+\frac1M\big)\right\}\cap\{\tau_k< M\}\right)\\[10pt]
&\le& \summ_{t\ge M}\P_{\mb\pi^k}\left(\left\{G^k_{\tau_k}>
\gamma\big(\eps+\frac1M\big)\right\}\cap\{\tau_k=t\}\right) +
\P_{\mb\pi^k}\left(\tau_k< M\right)\\[10pt]
&\le& \summ_{t\ge
M}\P_{\mb\pi^k}\left(G^k_t>\gamma\big(\eps+\frac1M\big)\right) +
\P_{\mb\pi^k}\left(\tau_k< M\right)
\\[10pt]
&\le& \summ_{t\ge M}2|\mc S|\ds\frac{\exp\left(-t\eps^2/2\right)}
{1-\exp\left(-\eps^2/2\right)} + \P_{\mb\pi^k}\left(\tau_k<
M\right)
\\[10pt]
&=& \ds\frac{2|\mc
S|}{\left(1-\exp\left(-\eps^2/2\right)\right)^2}
\exp\left(-M\eps^2/2\right) + \P_{\mb\pi^k}\left(\tau_k<
M\right)\,, \ea
$$
so that it follows from (\ref{probabilisticdivergence})
$$
\ba{rcl}
\limsup\limits_{k\in\N}\P_{\mb\pi^k}\left(G^k_{\tau_k}>\gamma\big(\eps+\frac1M\big)\right)
&\le& \ds\frac{2|\mc
S|}{\left(1-\exp\left(-\eps^2/2\right)\right)^2}
\exp\left(-M\eps^2/2\right)
+\limsup_{k\in\N}\P_{\mb\pi^k}\left(\tau_k\le M\right)\\[10pt]
&\le& \ds\frac{2|\mc S|}{\left(1-\exp\left(-\eps^2/4|\mc
S|^2\right)\right)^2} \exp\left(-M\eps^2/2\right)\,,\ea
$$
and by the arbitrariness of $M$ in $\N$ we get the claim. \qed

\setcounter{theorem}{11}
\begin{lemma}
In the previous setting, for every fixed $M$ in $\N$, we have
$$\lim\limits_{k\in\N}\P_{\Phi^k}\left(\tau_k\le
M\right)=0\,,\qquad\qquad
\lim_{k\in\N}\P_{\Phi^k}\left(T_k-\tau_k\le M\right)=0\,.
$$
\end{lemma}
\proof From Lemma \ref{lemmaPMAP} we have
$$P_{MAP}^{\Phi^k}(T_k)\ge
P_{MAP}^{\Phi^k}(\tau_k)\lambda^{T_k-\tau_k}\ge
\lambda\delta_k\lambda^{T_k-\tau_k}\,.$$ This implies that, for
every $M$ in $\N$,
$$
\ba{rcl} p_e\left(\Phi^k,T_k,\Psi^k\right)&=&
\E_{\Phi^k}\left[P_{MAP}^{\Phi^k}(T_k)\right]\\[5pt]
&\ge& \E_{\Phi^k}\left[P_{MAP}^{\Phi^k}(T_k)|T_k-\tau_k\le
M\right]\P_{\Phi^k}\left(T_k-\tau_k\le M\right)\\[5pt]
&\ge& \lambda\delta_k\lambda^{M}\P_{\Phi_k}\left(T_k-\tau_k\le
M\right)\,.\ea$$ It follows that
$$
\P_{\Phi^k}\left(T_k-\tau_k\le M\right)\le
\lambda^{-M-1}\frac{p_e\left(\Phi^k,T_k,\Psi^k\right)}{\delta_k}
\stackrel{k\ra\infty}{\longrightarrow}0\,.
$$

In order to show the first part of the claim,
suppose that $P_{MAP}^{\Phi^k}(\tau_k)\le\delta_k$. Then
$$\frac{|\mc W_k|-1}{|\mc W_k|}\lambda^{\tau_k}\le
P_{MAP}^{\Phi^k}\left(\tau_k\right)\le\delta_k\,.
$$
It follows that we have
$$
\ba{rcl} \P_{\Phi^k}\left(\tau_k\le M\right)&\le&
\P_{\Phi^k}\left(\left\{\tau_k\le M\right\}
\cap\left\{P_{MAP}^{\Phi^k}(\tau_k)\le\delta_k\right\}\right)
+\P_{\Phi^k}\left(P_{MAP}^{\Phi^k}(\tau_k)>\delta_k\right)\\[10pt]
&\le& \P_{\Phi^k}\left(\frac{|\mc W_k|-1}{|\mc W_k|}
\lambda^{M}\le\delta_k\right) +\P_{\Phi^k}\left(\tau_k=T_k\right)
\stackrel{k\ra\infty}{\longrightarrow}0\,. \ea
$$
\qed

\end{document}